\documentclass[journal,oneside,11pt,draftcls,onecolumn,a4paper]{IEEEtran}
\usepackage{amsfonts}
\usepackage{amssymb}
\usepackage{amsmath}
\usepackage[pdftex]{graphicx,graphics}

\ifCLASSINFOpdf
\else
\fi
\input{tcilatex}

\begin{document}

\title{A class of Weiss-Weinstein bounds and its relationship with the
Bobrovsky-Mayer-Wolf-Zaka\"{\i} bounds}
\author{Eric Chaumette, Alexandre Renaux and Mohammed Nabil El Korso \thanks{%
Eric Chaumette is with ISAE-SUPAERO, Universit\'{e} de Toulouse, 31055
Toulouse Cedex 4, France. Email: eric.chaumette@isae.fr}\thanks{%
Alexandre Renaux is with Universite Paris-Sud/LSS 3, Rue Joliot-Curie, 91192
Gif-sur-Yvette, France. Email: renaux@lss.supelec.fr}\thanks{%
Mohammed Nabil El Korso is with Laboratoire Energ\'{e}tique M\'{e}canique
Electromagn\'{e}tisme (LEME EA 4416), Universit\'{e} Paris Ouest Nanterre La
D\'{e}fense, IUT de Ville d'Avray, France. Email : m.elkorso@u-paris10.fr}}
\maketitle

\begin{abstract}
A fairly general class of Bayesian "large-error" lower bounds of the
Weiss-Weinstein family, essentially free from regularity conditions on the
probability density functions support, and for which a limiting form yields
a generalized Bayesian Cram\'{e}r-Rao bound (BCRB), is introduced. In a
large number of cases, the generalized BCRB appears to be the
Bobrovsky-Mayer-Wolf-Zakai bound (BMZB). Interestingly enough, a regularized
form of the Bobrovsky-Zakai bound (BZB), applicable when the support of the
prior is a constrained parameter set, is obtained. Modified Weiss-Weinstein
bound and BZB which limiting form is the BMZB are proposed, in expectation
of an increased tightness in the threshold region. Some of the proposed
results are exemplified with a reference problem in signal processing: the
Gaussian observation model with parameterized mean and uniform prior.
\end{abstract}


\section{Introduction}

Under the mean square error (MSE) criterion, the mean of the a posteriori
probability density function (pdf) of a random parameter, conditioned on the
observed data, is the optimal solution to the parameter estimation problem.
However, except for a few special cases, determining the posterior mean is
computationally prohibitive, and various approaches have been developed as
alternatives. It is therefore of interest to determine the degradation in
accuracy resulting from the use of suboptimal methods \cite{Simon}\cite%
{Sarkka}. Unfortunately again, the computation of the MSE of the conditional
mean estimator generally requires multiple integration, a computationally
intensive task \cite{Simon}\cite{Sarkka}. This has led to a large body of
work \cite{Van Trees - Bell}\cite{RFLRN08}\cite{Todros - Tabrikian - IT -
Bayesian} seeking to find both computationally tractable and tight Bayesian
lower bounds (BLBs) on the attainable MSE to which the performance of the
optimal estimator or any suboptimal estimation scheme can be compared.

Historically, computational tractability and ease of use seem to have been
the prominent qualities requested for a lower bound, as exemplified by the
Bayesian Cram\'{e}r-Rao bound (BCRB), the first Bayesian lower bound to be
derived \cite{Shutzenberger}\cite{Van Trees Part I}, and still the most
commonly used BLB. Nevertheless, it is now well known that the BCRB is an
optimistic bound in a non-linear estimation problem where the outliers
effect generally appears, leading to a characteristic behavior of estimators
MSE which exhibits three distinct regions of operation depending on the
number of (independent) observations and/or on the signal to noise ratio
(SNR) \cite{Van Trees - Bell}. More precisely, at high SNR and/or for a high
number of observations, i.e., in the asymptotic region, the outliers effect
can be neglected and the ultimate performance are generally described by the
BCRB. However, when the SNR and/or the number of observations decrease, the
outliers effect leads to a quick increase of the MSE: this is the so-called
threshold effect which is not predicted by the BCRB. Finally, at low SNR
and/or at low number of observations, the observations provide little
information, and the MSE is close to that obtained from the prior knowledge
about the problem yielding the no-information region.

Therefore after computational tractability, tightness and/or relaxation of
some regularity assumptions on the problem setting \cite{Ziv - Zakai}\cite%
{Bellini - Tartara}\cite{Bobrovsky - Zakai}\cite{Weiss - Weinstein 85} have
become the prominent qualities looked for a lower bound in non-linear
estimation problems. Indeed, from a practical point of view, the knowledge
of the particular value for which the threshold effect appears is a key
feature allowing to define estimators optimal operating area. This has led
to a large body of research based, so far, on two main families, i) the
Ziv-Zakai family (ZZF) resulting from the conversion of an estimation
bounding problem into one bounding binary hypothesis testing \cite{Ziv -
Zakai}\cite{Bellini - Tartara}\cite{Bell - 1997} and, ii) the
Weiss-Weinstein family (WWF), derived from a covariance inequality principle
\cite{Todros - Tabrikian - IT - Bayesian}\cite{Shutzenberger}\cite{Van Trees
Part I}\cite{Bobrovsky - Zakai}\cite{Weiss - Weinstein 85}\cite{Weiss -
Weinstein 88}\cite{BMWZ87}\cite{Reuven - Messer}\cite{Renaux AB}\cite{Bell -
Van Trees}. In each family, some bounds, generally called "large-error"
bounds (in contrast with "small-error" bounds such as the BCRB), can predict
the threshold effect \cite{Van Trees - Bell}.

In the present paper we focus on the Weiss-Weinstein family. The main
contribution of the paper is to introduce a fairly general class of
"large-error" bounds of the WWF essentially free from regularity conditions
and for which a limiting form yields a generalized BCRB. Indeed, within this
class of lower bounds, the supports of the joint and conditional pdfs must
only be a countable union of disjoint non empty intervals of $%
\mathbb{R}
$ (which naturally includes connected or disconnected subsets of $%
\mathbb{R}
$, bounded or unbounded intervals) and the bound-generating functions must
only have a finite second order moment. Additionally, we provide
(Propositions 1 and 2) some mild regularity conditions in order to obtain a
non trivial limiting form (non zero generalized BCRB) of the "large-error"
bound considered. In a large number of cases, this limiting form appears to
be the Bobrovsky-Mayer-Wolf-Zakai bound (BMZB) \cite{BMWZ87}. Therefore, the
proposed class of Bayesian lower bounds defines a wide range of Bayesian
estimation problems for which a non trivial generalized BCRB exists, which
is a key result from a practical viewpoint. Indeed, the computational cost
of large-error bounds is prohibitive in most applications when the number of
unknown parameters increases.\newline
Interestingly enough, the proposed class of lower bounds provides the
expression of all existing bounds of the WWF mentioned in \cite{RFLRN08} and
\cite{Todros - Tabrikian - IT - Bayesian} when the pdfs support is a
constrained parameter set, including a regularized form of the
Bobrovsky-Zakai bound (BZB) \cite{Bobrovsky - Zakai}. From a practical
viewpoint, it is another noticeable result, since the BZB is the easiest to
use "large-error" bound, but was believed to be inapplicable in that case
\cite[Section II]{Bobrovsky - Zakai}\cite[p682]{Weiss - Weinstein 85}\cite[%
p340]{Weiss - Weinstein 88}\cite[p39]{Van Trees - Bell}. \newline
Last, as a by-product, since the BMZB may provide a tighter bound than the
historical BCRB in the asymptotic region \cite{BMWZ87}\cite[p36]{Van Trees -
Bell}, it would seem sensible to introduce modified Weiss-Weinstein bound
(WWB) and BZB which limiting form is the BMZB, in expectation of an
increased tightness in the threshold region as well.

Some of the proposed results are exemplified with a reference problem in
signal processing: the Gaussian observation model with parameterized mean
depending on a random parameter with uniform prior. For numerical
evaluations, we focus on the estimation of a single tone.

For sake of legibility, we only discuss in details the case of a single
random parameter. Extension of the proposed results to a vector of
parameters can be done by resorting to the covariance matrix inequality as
shown in \cite[p341]{Weiss - Weinstein 88}\cite[p1429]{BMWZ87}.

\section{A new class of Bayesian lower bounds of the Weiss-Weinstein family}

Throughout the present paper scalars, vectors and matrices are represented,
respectively, by italic (as in $a$ or $A$), bold lowercase (as in $\mathbf{a}
$) and bold uppercase (as in $\mathbf{A}$) characters. The $n$-th row and $m$%
-th column element of the matrix $\mathbf{A}$ is denoted by $\left\{ \mathbf{%
A}\right\} _{n,m}$, whereas, $\left\{ \mathbf{a}\right\} _{n}$ represents
the $n$-th coordinate of the column vector $\mathbf{a}$. The real and
imaginary part of $A$, are denoted, respectively, by $\func{Re}\left\{
A\right\} $ and $\func{Im}\left\{ A\right\} $. The transpose, transpose
conjugate operator are indicated, respectively, by $.^{T}$ and $.^{H}$. The
identity matrix of size $M$ is denoted by $\mathbf{I}_{M}$. For any given
two matrices\textit{\ }$\mathbf{A}$\textit{\ }and\textit{\ }$\mathbf{B}$, $%
\mathbf{A}\succeq \mathbf{B}$\textit{\ }means that\textit{\ }$\mathbf{A}-%
\mathbf{B}$\textit{\ }is positive semi-definite matrix. $E\left[ .\right] $
denotes the expectation operator and $1_{A}\left( \mathbf{x}\right) $ is the
indicator function of subset $A$ of $\mathbb{R}^{N}$.

\subsection{Definitions and Assumptions}

Throughout the present paper:

\begin{itemize}
\item $\mathbf{x}$ denotes a $N$-dimensional complex random observation
vector belonging to the observation space $\mathcal{X}\subset \mathbb{C}^{N}$%
.

\item $\theta $ denotes a real random parameter belonging to the parameter
space $\Theta \subset \mathbb{R}$.

\item $\mathcal{S}_{\mathcal{X},\Theta }\subset \mathbb{C}^{N}\times \mathbb{%
R}$ denotes the support of the the joint pdf $p\left( \mathbf{x},\theta
\right) $ of $\mathbf{x}$ and $\theta $ such that $\mathcal{S}_{\mathcal{X}%
,\Theta }=\left\{ \left( \mathbf{x}^{T},\theta \right) ^{T}\in \mathbb{C}%
^{N}\times \mathbb{R}\text{ : }p\left( \mathbf{x},\theta \right) >0\right\} $%
.

\item $\mathcal{S}_{\Theta }\subset \mathbb{R}$ denotes the support of the
prior pdf of $\theta $ denoted $p\left( \theta \right) $, i.e., $\mathcal{S}%
_{\Theta }=\left\{ \theta \in \mathbb{R}:p\left( \theta \right) >0\right\} $.

\item $\mathcal{S}_{\mathcal{X}}\subset \mathbb{C}^{N}$ denotes the support
of the marginal pdf of $\mathbf{x}$ denoted $p\left( \mathbf{x}\right) $,
i.e., $\mathcal{S}_{\mathcal{X}}=\left\{ \mathbf{x}\in \mathbb{C}%
^{N}:p\left( \mathbf{x}\right) >0\right\} $.

\item Furthermore, $\forall \mathbf{x}\in \mathcal{S}_{\mathcal{X}}$, let us
denote $\mathcal{S}_{\Theta |\mathbf{x}}=\left\{ \theta \in \mathbb{R}%
:p\left( \mathbf{x},\theta \right) >0\right\} $ and $\forall \theta \in
\mathcal{S}_{\Theta }$, $\mathcal{S}_{\mathcal{X}|\theta }=\left\{ \mathbf{x}%
\in \mathbb{C}^{N}:p\left( \mathbf{x},\theta \right) >0\right\} $. Then:%
\begin{equation*}
p\left( \theta \right) =\dint\limits_{\mathbb{C}^{N}}p\left( \mathbf{x}%
,\theta \right) 1_{\mathcal{S}_{\mathcal{X}|\theta }}\left( \mathbf{x}%
\right) d\mathbf{x}=\dint\limits_{\mathcal{S}_{\mathcal{X}|\theta }}p\left(
\mathbf{x},\theta \right) d\mathbf{x,\quad }p\left( \mathbf{x}\right)
=\dint\limits_{\mathbb{R}}p\left( \mathbf{x},\theta \right) 1_{\mathcal{S}_{%
\mathbf{\Theta }|\mathbf{x}}}\left( \mathbf{x}\right) d\theta =\dint\limits_{%
\mathcal{S}_{\Theta |\mathbf{x}}}p\left( \mathbf{x},\theta \right) d\theta .
\end{equation*}
\end{itemize}

Thus, for a given function $f:\mathcal{X}\times \Theta \rightarrow \mathbb{R}
$, deterministic, unknown and measurable function, one has:
\begin{eqnarray*}
E_{\mathbf{x},\theta }\left[ f\left( \mathbf{x},\theta \right) \right]
&=&\dint\limits_{\mathbb{C}^{N}}\dint\limits_{\mathbb{R}}f\left( \mathbf{x}%
,\theta \right) p\left( \mathbf{x},\theta \right) 1_{\mathcal{S}_{\mathcal{X}%
,\Theta }}(\mathbf{x},\theta )d\mathbf{x}d\theta =\dint\limits_{\mathcal{S}_{%
\mathcal{X}}}\dint\limits_{\mathcal{S}_{\Theta |\mathbf{x}}}f\left( \mathbf{x%
},\theta \right) p\left( \mathbf{x},\theta \right) d\mathbf{x}d\theta , \\
E_{\mathbf{x}|\theta }\left[ f\left( \mathbf{x},\theta \right) \right]
&=&\dint\limits_{\mathbb{C}^{N}}f\left( \mathbf{x},\theta \right) p\left(
\mathbf{x}|\theta \right) 1_{\mathcal{S}_{\mathcal{X}|\theta }}(\mathbf{x})d%
\mathbf{x}=\dint\limits_{\mathcal{S}_{\mathcal{X}|\theta }}f\left( \mathbf{x}%
,{\theta }\right) p\left( \mathbf{x}|\theta \right) d\mathbf{x,} \\
E_{{\theta |\mathbf{x}}}\left[ f\left( \mathbf{x},\theta \right) \right]
&=&\dint\limits_{\mathbb{R}}f\left( \mathbf{x},\theta \right) p\left( {%
\theta |\mathbf{x}}\right) 1_{\mathcal{S}_{\Theta |\mathbf{x}}}(\theta
)d\theta =\dint\limits_{\mathcal{S}_{\Theta |\mathbf{x}}}f\left( \mathbf{x}%
,\theta \right) p\left( {\theta |\mathbf{x}}\right) d\theta , \\
E_{\mathbf{x}}\left[ f\left( \mathbf{x},\theta \right) \right]
&=&\dint\limits_{\mathbb{C}^{N}}f\left( \mathbf{x},\theta \right) p\left(
\mathbf{x}\right) 1_{\mathcal{S}_{\mathcal{X}}}(\mathbf{x})d\mathbf{x}%
=\dint\limits_{\mathcal{S}_{\mathcal{X}}}f\left( \mathbf{x},\theta \right)
p\left( \mathbf{x}\right) d\mathbf{x,} \\
E_{\theta }\left[ f\left( \mathbf{x},\theta \right) \right] &=&\dint\limits_{%
\mathbb{R}}f\left( \mathbf{x},\theta \right) p\left( \theta \right) 1_{%
\mathcal{S}_{\Theta }}(\theta )d\theta =\dint\limits_{\mathcal{S}_{\Theta
}}f\left( \mathbf{x},\theta \right) p\left( {\theta }\right) d\theta ,
\end{eqnarray*}%
\begin{eqnarray*}
E_{\mathbf{x},\theta }\left[ f\left( \mathbf{x},\theta \right) \right]
&=&\dint\limits_{\mathcal{S}_{\mathcal{X}}}\left( \dint\limits_{\mathcal{S}%
_{\Theta |\mathbf{x}}}f\left( \mathbf{x},\theta \right) \frac{p\left(
\mathbf{x},\theta \right) }{p\left( \mathbf{x}\right) }d\theta \right)
p\left( \mathbf{x}\right) d\mathbf{x}=E_{\mathbf{x}}\left[ E_{\theta |%
\mathbf{x}}\left[ f\left( \mathbf{x},\theta \right) \right] \right] , \\
E_{\mathbf{x},\theta }\left[ f\left( \mathbf{x},\theta \right) \right]
&=&\dint\limits_{\mathcal{S}_{\Theta }}\left( \dint\limits_{\mathcal{S}_{%
\mathcal{X}|\theta }}f\left( \mathbf{x},\theta \right) \frac{p\left( \mathbf{%
x},\theta \right) }{p\left( \theta \right) }d\mathbf{x}\right) p\left(
\theta \right) d\theta =E_{\theta }\left[ E_{\mathbf{x}|\theta }\left[
f\left( \mathbf{x},\theta \right) \right] \right] .
\end{eqnarray*}%
Additionally, we assume that:

\begin{itemize}
\item \textbf{A1)} $g(\theta ):\mathbb{R}\rightarrow \mathbb{R}$, $g(.)\in
\mathcal{L}_{2}(\mathcal{S}_{\Theta |\mathbf{x}})$, $\forall \mathbf{x}\in
\mathcal{S}_{\mathcal{X}}$, is the deterministic, known, measurable function
to be estimated, where $\mathcal{L}_{2}(\mathcal{S}_{\Theta |\mathbf{x}})$
denotes the space of square integrable functions w.r.t. $p(\theta |\mathbf{x}%
)$, i.e., $E_{\theta |\mathbf{x}}[g(\theta )^{2}]<\infty $.

\item \textbf{A2)} $\hat{g}(\mathbf{x}):\mathcal{X}\rightarrow \mathbb{R}$, $%
\hat{g}(\mathbf{.})\in \mathcal{L}_{2}(\mathcal{S}_{\mathcal{X}})$, denotes
any deterministic, known, measurable estimator of $g(\theta )$, where $%
\mathcal{L}_{2}(\mathcal{S}_{\mathcal{X}})$ denotes the space of square
integrable functions w.r.t. $p(\mathbf{x})$, i.e., $E_{\mathbf{x}}[\hat{g}(%
\mathbf{x})^{2}]<\infty $.

\item \textbf{A3) }$\psi \left( \mathbf{x},\theta \right) :\mathcal{X}\times
\mathbb{R}\rightarrow \mathbb{R}$, $\psi \left( \mathbf{.}\right) \in
\mathcal{L}_{2}(\mathcal{S}_{\mathcal{X},\Theta })$, denotes a
deterministic, known, measurable function, where $\mathcal{L}_{2}(\mathcal{S}%
_{\mathcal{X},\Theta })$ denotes the space of square integrable functions
w.r.t. $p(\mathbf{x},\theta )$, i.e., $E_{\mathbf{x},\theta }[\psi (\mathbf{x%
},\theta )^{2}]<\infty $, and satisfying $0<E_{\mathbf{x},\theta }[\psi (%
\mathbf{x},\theta )^{2}]$.
\end{itemize}

\subsection{Background on covariance inequality}

Under the assumptions A1), A2) and A3), the Cauchy-Schwartz inequality
states that:
\begin{subequations}
\begin{equation}
E_{\mathbf{x},\theta }\left[ \left( \hat{g}(\mathbf{x})-g\left( \theta
\right) \right) \psi \left( \mathbf{x},\theta \right) \right] ^{2}\leq E_{%
\mathbf{x},\theta }\left[ \left( \hat{g}(\mathbf{x})-g\left( \theta \right)
\right) ^{2}\right] E_{\mathbf{x},\theta }\left[ \psi \left( \mathbf{x}%
,\theta \right) ^{2}\right] .
\end{equation}%
Therefore:
\begin{equation}
E_{\mathbf{x},\theta }\left[ \left( \hat{g}(\mathbf{x})-g\left( \theta
\right) \right) ^{2}\right] \geq \frac{E_{\mathbf{x},\theta }\left[ \left(
\hat{g}(\mathbf{x})-g\left( \theta \right) \right) \psi \left( \mathbf{x}%
,\theta \right) \right] ^{2}}{E_{\mathbf{x},\theta }\left[ \psi \left(
\mathbf{x},\theta \right) ^{2}\right] }=\frac{\left( E_{\mathbf{x},\theta }%
\left[ \hat{g}(\mathbf{x})\psi \left( \mathbf{x},\theta \right) \right] -E_{%
\mathbf{x},\theta }\left[ g\left( \theta \right) \psi \left( \mathbf{x}%
,\theta \right) \right] \right) ^{2}}{E_{\mathbf{x},\theta }\left[ \psi
\left( \mathbf{x},\theta \right) ^{2}\right] }.
\label{covariance inequality - g(x)}
\end{equation}%
A necessary condition on $\psi \left( \mathbf{x},\theta \right) $ in order
to obtain a lower bound on the MSE of $\hat{g}(\mathbf{x})$, i.e., an
expression independent from the estimator $\hat{g}(\mathbf{x})$ in the
right-hand side of (\ref{covariance inequality - g(x)}), is to satisfy \cite%
{Weiss - Weinstein 88}:
\end{subequations}
\begin{subequations}
\begin{equation}
E_{\mathbf{x},\theta }\left[ \hat{g}(\mathbf{x})\psi \left( \mathbf{x}%
,\theta \right) \right] =0.  \label{Cond Suffisante Generale - v0}
\end{equation}%
As $\hat{g}\left( \mathbf{x}\right) $ is $\theta $ independent, thus, (\ref%
{Cond Suffisante Generale - v0}) can be rewritten as:
\begin{equation}
E_{\mathbf{x},\theta }\left[ \hat{g}(\mathbf{x})\psi \left( \mathbf{x}%
,\theta \right) \right] =E_{\mathbf{x}}\left[ E_{{\theta |\mathbf{x}}}\left[
\hat{g}(\mathbf{x})\psi \left( \mathbf{x},\theta \right) \right] \right] =E_{%
\mathbf{x}}\left[ \hat{g}\left( \mathbf{x}\right) E_{\theta |\mathbf{x}}%
\left[ \psi \left( \mathbf{x},\theta \right) \right] \right] .
\end{equation}%
Consequently, a sufficient condition for a judicious choice of $\psi \left(
\mathbf{x},\theta \right) $ is simply \cite{Weiss - Weinstein 88}:
\begin{equation}
E_{\theta |\mathbf{x}}\left[ \psi \left( \mathbf{x},\theta \right) \right]
=0.  \label{Cond Suffisante Generale}
\end{equation}%
Finally, a non trivial bound is obtained from (\ref{covariance inequality -
g(x)}) for the family of functions $\psi \left( \mathbf{x},\theta \right) $
satisfying both (\ref{Cond Suffisante Generale}) and $E_{\mathbf{x},\theta }%
\left[ g\left( \theta \right) \psi \left( \mathbf{x},\theta \right) \right]
\neq 0$, yielding the Weiss-Weinstein family of Bayesian lower bounds \cite%
{Weiss - Weinstein 88} given by:
\end{subequations}
\begin{equation}
E_{\mathbf{x},\theta }\left[ \left( \hat{g}(\mathbf{x})-g\left( \theta
\right) \right) ^{2}\right] \geq \frac{E_{\mathbf{x},\theta }\left[ g\left(
\theta \right) \psi \left( \mathbf{x},\theta \right) \right] ^{2}}{E_{%
\mathbf{x},\theta }\left[ \psi \left( \mathbf{x},\theta \right) ^{2}\right] }%
.  \label{covariance inequality}
\end{equation}

\subsection{Proposed class of Bayesian lower bounds}

Let us consider a function $q\left( \mathbf{x},\theta \right) :\mathcal{X}%
\times \mathbb{R}\rightarrow \mathbb{R}$. Thus, one can notice that, since $%
p\left( \theta |\mathbf{x}\right) =p\left( \theta |\mathbf{x}\right) 1_{%
\mathcal{S}_{\Theta |\mathbf{x}}}\left( \theta \right) $, then, $\forall
\mathbf{x}\in \mathcal{S}_{\mathcal{X}}$:
\begin{subequations}
\begin{align}
\dint\limits_{\mathcal{S}_{\Theta |\mathbf{x}}}q\left( \mathbf{x},\theta
+h\right) p\left( \theta +h|\mathbf{x}\right) 1_{\mathcal{S}_{\Theta |%
\mathbf{x}}}\left( \theta \right) d\theta & =\dint\limits_{\mathbb{R}%
}q\left( \mathbf{x},\theta +h\right) p\left( \theta +h|\mathbf{x}\right) 1_{%
\mathcal{S}_{\Theta |\mathbf{x}}}\left( \theta +h\right) 1_{\mathcal{S}%
_{\Theta |\mathbf{x}}}\left( \theta \right) d\theta  \\
& =\dint\limits_{\mathbb{R}}q\left( \mathbf{x},\theta \right) p\left( \theta
|\mathbf{x}\right) 1_{\mathcal{S}_{\Theta |\mathbf{x}}}\left( \theta \right)
1_{\mathcal{S}_{\Theta |\mathbf{x}}}\left( \theta -h\right) d\theta  \\
& =\dint\limits_{\mathcal{S}_{\Theta |\mathbf{x}}}q\left( \mathbf{x},\theta
\right) p\left( \theta |\mathbf{x}\right) 1_{\mathcal{S}_{\Theta |\mathbf{x}%
}}\left( \theta -h\right) 1_{\mathcal{S}_{\Theta |\mathbf{x}}}\left( \theta
\right) d\theta
\end{align}%
leading to:
\end{subequations}
\begin{equation}
\dint\limits_{\mathcal{S}_{\Theta |\mathbf{x}}}q\left( \mathbf{x},\theta
+h\right) p\left( \theta +h|\mathbf{x}\right) 1_{\mathcal{S}_{\Theta |%
\mathbf{x}}}\left( \theta \right) d\theta -\dint\limits_{\mathcal{S}_{\Theta
|\mathbf{x}}}q\left( \mathbf{x},\theta \right) p\left( \theta |\mathbf{x}%
\right) 1_{\mathcal{S}_{\Theta |\mathbf{x}}}\left( \theta -h\right) 1_{%
\mathcal{S}_{\Theta |\mathbf{x}}}\left( \theta \right) d\theta =0.
\end{equation}%
Consequently, in order to fulfill (\ref{Cond Suffisante Generale}), we
propose to use the following class of bound-generating functions:
\begin{equation}
\psi _{q}^{h}\left( \mathbf{x},\theta \right) =%
\begin{cases}
\left( \frac{p\left( \theta +h|\mathbf{x}\right) }{p\left( \theta |\mathbf{x}%
\right) }q\left( \mathbf{x},\theta +h\right) -q\left( \mathbf{x},\theta
\right) 1_{\mathcal{S}_{\Theta |\mathbf{x}}}\left( \theta -h\right) \right)
1_{\mathcal{S}_{\Theta |\mathbf{x}}}\left( \theta \right) , & \text{if }%
\left( \mathbf{x},\theta \right) \in \mathcal{S}_{\mathcal{X},\Theta } \\
0, & \text{otherwise }%
\end{cases}%
,  \label{psi(h,f) - WWF}
\end{equation}%
for which the choice of the function $q(\mathbf{.})$ is only subject to: $%
0<E_{\mathbf{x},\theta }\left[ \psi _{q}^{h}\left( \mathbf{x},\theta \right)
^{2}\right] <\infty $.\newline
Now, we can derive the right-hand side of (\ref{covariance inequality}). As:
\begin{multline}
E_{\theta |\mathbf{x}}\left[ g\left( \theta \right) \psi _{q}^{h}\left(
\mathbf{x},\theta \right) \right] =\dint\limits_{\mathcal{S}_{\Theta {|%
\mathbf{x}}}}g\left( \theta \right) q\left( \mathbf{x},\theta +h\right)
p\left( \theta +h|\mathbf{x}\right) 1_{\mathcal{S}_{\Theta |\mathbf{x}%
}}\left( \theta \right) d\theta  \\
-\dint\limits_{\mathcal{S}_{\Theta {|\mathbf{x}}}}g\left( \theta \right)
q\left( \mathbf{x},\theta \right) 1_{\mathcal{S}_{\Theta |\mathbf{x}}}\left(
\theta -h\right) p\left( \theta |\mathbf{x}\right) d\theta ,
\end{multline}%
and the first integral of the above equation can be written as:
\begin{subequations}
\begin{equation}
\dint\limits_{\mathcal{S}_{\Theta {|\mathbf{x}}}}g\left( \theta \right)
q\left( \mathbf{x},\theta +h\right) p\left( \theta +h|\mathbf{x}\right) 1_{%
\mathcal{S}_{\Theta |\mathbf{x}}}\left( \theta \right) d\theta
=\dint\limits_{\mathbb{R}}g\left( \theta \right) q\left( \mathbf{x},\theta
+h\right) p\left( \theta +h|\mathbf{x}\right) 1_{\mathcal{S}_{\Theta |%
\mathbf{x}}}\left( \theta +h\right) 1_{\mathcal{S}_{\Theta |\mathbf{x}%
}}\left( \theta \right) d\theta ,\qquad
\end{equation}%
\begin{eqnarray}
\qquad \qquad \qquad \qquad \qquad \qquad \qquad \qquad \quad
&=&\dint\limits_{\mathbb{R}}g\left( \theta -h\right) q\left( \mathbf{x}%
,\theta \right) p\left( \theta |\mathbf{x}\right) 1_{\mathcal{S}_{\Theta |%
\mathbf{x}}}\left( \theta \right) 1_{\mathcal{S}_{\Theta |\mathbf{x}}}\left(
\theta -h\right) d\theta ,\qquad \quad  \\
&=&\dint\limits_{\mathcal{S}_{\Theta |\mathbf{x}}}g\left( \theta -h\right)
q\left( \mathbf{x},\theta \right) 1_{\mathcal{S}_{\Theta |\mathbf{x}}}\left(
\theta -h\right) p\left( \theta |\mathbf{x}\right) d\theta ,
\end{eqnarray}%
therefore:
\end{subequations}
\begin{subequations}
\begin{eqnarray}
E_{\theta |\mathbf{x}}\left[ g\left( \theta \right) \psi _{q}^{h}\left(
\mathbf{x},\theta \right) \right]  &=&\dint\limits_{\mathcal{S}_{\Theta |%
\mathbf{x}}}\left( g\left( \theta -h\right) -g\left( \theta \right) \right)
q\left( \mathbf{x},\theta \right) 1_{\mathcal{S}_{\Theta |\mathbf{x}}}\left(
\theta -h\right) p\left( \theta |\mathbf{x}\right) d\theta  \\
&=&E_{\theta |\mathbf{x}}\left[ \left( g\left( \theta -h\right) -g\left(
\theta \right) \right) q\left( \mathbf{x},\theta \right) 1_{\mathcal{S}%
_{\Theta |\mathbf{x}}}\left( \theta -h\right) 1_{\mathcal{S}_{\Theta |%
\mathbf{x}}}\left( \theta \right) \right]
\end{eqnarray}%
Finally, the proposed class of BLBs is given by:
\end{subequations}
\begin{equation}
\mathrm{BLB}_{q}^{h}\left( g\left( \theta \right) \right) =\frac{E_{\mathbf{x%
},\theta }\left[ \left( g\left( \theta -h\right) -g\left( \theta \right)
\right) q\left( \mathbf{x},\theta \right) 1_{\mathcal{S}_{\Theta |\mathbf{x}%
}}\left( \theta -h\right) 1_{\mathcal{S}_{\Theta |\mathbf{x}}}\left( \theta
\right) \right] ^{2}}{E_{\mathbf{x},\theta }\left[ \left( q\left( \mathbf{x}%
,\theta +h\right) \frac{p\left( \theta +h|\mathbf{x}\right) }{p\left( \theta
|\mathbf{x}\right) }-q\left( \mathbf{x},\theta \right) 1_{\mathcal{S}%
_{\Theta |\mathbf{x}}}\left( \theta -h\right) \right) ^{2}1_{\mathcal{S}%
_{\Theta |\mathbf{x}}}\left( \theta \right) \right] },  \label{BLB}
\end{equation}%
and tighter BLBs can be obtained as:
\begin{equation}
\underset{q_{l}\left( .\right) ,1\leq l\leq L,h\in
\mathbb{R}
:~\psi _{q_{l}}^{h}\left( .\right) \in \mathcal{L}_{2}(\mathcal{S}_{\mathcal{%
X},\Theta })}{\sup }\left\{ \mathrm{BLB}_{q_{l}}^{h}\left( g\left( \theta
\right) \right) \right\} .  \label{BLB tighther}
\end{equation}%
Let us recall that Bayesian lower bounds are actually \emph{posterior} lower
bounds, i.e. lower bounding the MSE of the \emph{posterior} mean $E_{\theta |%
\mathbf{x}}\left[ g\left( \theta \right) \right] $. However as:
\begin{subequations}
\begin{equation}
\frac{p\left( \theta +h|\mathbf{x}\right) }{p\left( \theta |\mathbf{x}%
\right) }=\frac{p\left( \mathbf{x,}\theta +h\right) }{p\left( \mathbf{x,}%
\theta \right) },\forall \left( \mathbf{x},\theta \right) \in \mathcal{S}_{%
\mathcal{X},\Theta },
\end{equation}%
we also resort to the alternative form of (\ref{BLB}):%
\begin{equation}
\mathrm{BLB}_{q}^{h}\left( g\left( \theta \right) \right) =\frac{E_{\mathbf{x%
},\theta }\left[ \left( g\left( \theta -h\right) -g\left( \theta \right)
\right) q\left( \mathbf{x},\theta \right) 1_{\mathcal{S}_{\Theta |\mathbf{x}%
}}\left( \theta -h\right) 1_{\mathcal{S}_{\Theta |\mathbf{x}}}\left( \theta
\right) \right] ^{2}}{E_{\mathbf{x},\theta }\left[ \left( q\left( \mathbf{x}%
,\theta +h\right) \frac{p\left( \mathbf{x,}\theta +h\right) }{p\left(
\mathbf{x,}\theta \right) }-q\left( \mathbf{x},\theta \right) 1_{\mathcal{S}%
_{\Theta |\mathbf{x}}}\left( \theta -h\right) \right) ^{2}1_{\mathcal{S}%
_{\Theta |\mathbf{x}}}\left( \theta \right) \right] }.
\end{equation}%
\newline

\section{A new class of BCRBs and its relationship with the BMZBs}

From the literature \cite{Weiss - Weinstein 88}\cite{Reuven - Messer}\cite[%
p39]{Van Trees - Bell}, the historical BCRB \cite{Van Trees Part I} is given
as the limiting form of the BZB where $\mathcal{S}_{\Theta |\mathbf{x}}=%
\mathbb{R}
$, that is:
\end{subequations}
\begin{equation}
\mathrm{BCRB}\left( g\left( \theta \right) \right) =\underset{h\rightarrow 0}%
{\lim }\frac{E_{\mathbf{x},\theta }\left[ g\left( \theta \right) \frac{1}{h}%
\left( \frac{p\left( \theta +h|\mathbf{x}\right) }{p\left( \theta |\mathbf{x}%
\right) }-1\right) \right] ^{2}}{E_{\mathbf{x},\theta }\left[ \left( \frac{1%
}{h}\left( \frac{p\left( \theta +h|\mathbf{x}\right) }{p\left( \theta |%
\mathbf{x}\right) }-1\right) \right) ^{2}\right] }=\frac{E_{\mathbf{x}%
,\theta }\left[ \frac{dg\left( \theta \right) }{d\theta }\right] ^{2}}{E_{%
\mathbf{x},\theta }\left[ \left( \frac{\partial \ln p\left( \theta |\mathbf{x%
}\right) }{\partial \theta }\right) ^{2}\right] }.
\end{equation}%
Mutatis mutandis, we can use this definition for every function $q\left(
\mathbf{x},\theta \right) $ in order to define a generalized BCRB as
follows:
\begin{equation}
\mathrm{BCRB}_{q}\left( g\left( \theta \right) \right) =\max \left\{
\lim_{h\rightarrow 0^{+}}\frac{E_{\mathbf{x},\theta }\left[ g\left( \theta
\right) \frac{1}{h}\psi _{q}^{h}\left( \mathbf{x},\theta \right) \right] ^{2}%
}{E_{\mathbf{x},\theta }\left[ \left( \frac{1}{h}\psi _{q}^{h}\left( \mathbf{%
x},\theta \right) \right) ^{2}\right] },\lim_{h\rightarrow 0^{-}}\frac{E_{%
\mathbf{x},\theta }\left[ g\left( \theta \right) \frac{1}{h}\psi
_{q}^{h}\left( \mathbf{x},\theta \right) \right] ^{2}}{E_{\mathbf{x},\theta }%
\left[ \left( \frac{1}{h}\psi _{q}^{h}\left( \mathbf{x},\theta \right)
\right) ^{2}\right] }\right\} .  \label{GBCRB}
\end{equation}%
Interestingly enough, under the assumptions A1), A2) and A3),\textbf{\ }any
"large-error" bounds of the proposed class, i.e. $\mathrm{BLB}_{q}^{h}\left(
g\left( \theta \right) \right) $ (\ref{BLB}), admits a finite limiting form $%
\mathrm{BCRB}_{q}\left( g\left( \theta \right) \right) $ (\ref{GBCRB}).
Moreover, under some mild regularity conditions (see Propositions 1 and 2
below), the generalized BCRB is non zero, and in a large number of cases,
this limiting form appears to be the BMZB \cite{BMWZ87}. Therefore, the
proposed class of BLBs defines a wide range of Bayesian estimation problems
for which a non trivial BCRB exists, which is a key result from a practical
viewpoint. Indeed, the computational cost of large-error bounds is
prohibitive in most applications when the number of unknown parameters
increases \cite{RFLRN08}\cite{Todros - Tabrikian - IT - Bayesian}.

\subsection{Case where $\mathcal{S}_{\Theta |\mathbf{x}}$ is an interval of $%
\mathbb{R}
$}

Then we can state the following\smallskip \newline
\textbf{Proposition 1} : If $\forall \mathbf{x}\in \mathcal{S}_{\mathcal{X}}$%
:\newline
$\bullet ~\mathcal{S}_{\Theta |\mathbf{x}}$ is an interval of $%
\mathbb{R}
$ with endpoints $a_{\mathbf{x}},b_{\mathbf{x}}\in \left[ -\infty ,+\infty %
\right] $, $a_{\mathbf{x}}<b_{\mathbf{x}}$,\newline
$\bullet ~q\left( \mathbf{x},\theta \right) $ admits a finite limit at
endpoints,\newline
$\bullet ~g\left( \theta \right) $ is piecewise $\mathcal{C}^{1}$ w.r.t. $%
\theta $ over $\mathcal{S}_{\Theta |\mathbf{x}}$,\newline
$\bullet ~t\left( \mathbf{x},\theta \right) \triangleq q\left( \mathbf{x}%
,\theta \right) p\left( \theta |\mathbf{x}\right) $ is piecewise $\mathcal{C}%
^{1}$ w.r.t. $\theta $ over $\mathcal{S}_{\Theta |\mathbf{x}}$ and such as $%
\frac{\partial t\left( \mathbf{x},\theta \right) }{\partial \theta }$ admits
a finite limit at endpoints,\newline
${\small \bullet }~u\left( \mathbf{x},\theta \right) \triangleq q\left(
\mathbf{x},\theta \right) t\left( \mathbf{x},\theta \right) $ is $\mathcal{C}%
^{2}$ w.r.t. $\theta $ at the vicinity of endpoints and such as $u\left(
\mathbf{x},\theta \right) ,\frac{\partial u\left( \mathbf{x},\theta \right)
}{\partial \theta }$ and $\frac{\partial ^{2}u\left( \mathbf{x},\theta
\right) }{\partial ^{2}\theta }$ admit a finite limit at endpoints,\newline
then a necessary and sufficient condition in order to obtain a non trivial $%
\mathrm{BCRB}_{q}$ bound (\ref{GBCRB}) is:
\begin{subequations}
\begin{equation}
\lim\limits_{\theta \rightarrow a_{\mathbf{x}}}t\left( \mathbf{x},\theta
\right) =0=\lim\limits_{\theta \rightarrow b_{\mathbf{x}}}t\left( \mathbf{x}%
,\theta \right) ,  \label{cond BCRB non trivial - connected subset}
\end{equation}%
which leads to:
\begin{equation}
\mathrm{BCRB}_{q}\left( g\left( \theta \right) \right) =\frac{E_{\mathbf{x}%
,\theta }\left[ \frac{dg\left( \theta \right) }{d\theta }q\left( \mathbf{x}%
,\theta \right) \right] ^{2}}{E_{\mathbf{x},\theta }\left[ \left( \frac{%
\frac{\partial t\left( \mathbf{x},\theta \right) }{\partial \theta }}{%
p\left( \theta |\mathbf{x}\right) }\right) ^{2}\right] +\min \left\{
\begin{array}{l}
E_{\mathbf{x}}\left[ \frac{5}{2}\lim\limits_{\theta \rightarrow a_{\mathbf{x}%
}}v\left( \mathbf{x},\theta \right) -\frac{1}{2}\lim\limits_{\theta
\rightarrow b_{\mathbf{x}}}v\left( \mathbf{x},\theta \right) \right]
,\smallskip  \\
E_{\mathbf{x}}\left[ \frac{1}{2}\lim\limits_{\theta \rightarrow a_{\mathbf{x}%
}}v\left( \mathbf{x},\theta \right) -\frac{5}{2}\lim\limits_{\theta
\rightarrow b_{\mathbf{x}}}v\left( \mathbf{x},\theta \right) \right]
\end{array}%
\right\} }  \label{BCRB - connected subset}
\end{equation}%
where $v\left( \mathbf{x},\theta \right) =q\left( \mathbf{x},\theta \right)
^{2}\frac{\partial p\left( \theta |\mathbf{x}\right) }{\partial \theta }$%
.\medskip \newline
\textit{Proof:} see Appendix \ref{A: Case of bounded intervals} and Appendix %
\ref{A: Case of unbounded intervals}.\medskip \newline
In order to obtain a tight $\mathrm{BCRB}_{q}\left( g\left( \theta \right)
\right) $, it seems judicious to choose $q\left( \mathbf{x},\theta \right) $
such that:
\end{subequations}
\begin{subequations}
\begin{equation}
\forall \mathbf{x}\in \mathcal{S}_{\mathcal{X}}:\lim\limits_{\theta
\rightarrow a_{\mathbf{x}}}q\left( \mathbf{x},\theta \right)
=0=\lim\limits_{\theta \rightarrow b_{\mathbf{x}}}q\left( \mathbf{x},\theta
\right) .  \label{cond 1}
\end{equation}%
Indeed, then (\ref{BCRB - connected subset}) reduces to:
\begin{equation}
\mathrm{BCRB}_{q}\left( g\left( \theta \right) \right) =\frac{E_{\mathbf{x}%
,\theta }\left[ \frac{dg\left( \theta \right) }{d\theta }q\left( \mathbf{x}%
,\theta \right) \right] ^{2}}{E_{\mathbf{x},\theta }\left[ \left( \frac{%
\frac{\partial t\left( \mathbf{x},\theta \right) }{\partial \theta }}{%
p\left( \theta |\mathbf{x}\right) }\right) ^{2}\right] }=\mathrm{BMZB}%
_{q}\left( g\left( \theta \right) \right) ,  \label{BMWZB}
\end{equation}%
where $\mathrm{BMZB}_{q}\left( g\left( \theta \right) \right) $ stands for
the BMZB \cite[(24)]{BMWZ87}.\smallskip \newline
Note that:\smallskip \newline
$\bullet $ the above condition (\ref{cond BCRB non trivial - connected
subset}) is not explicitly given in the original paper of \cite[\S 4]{BMWZ87}
nor in \cite[p35]{Van Trees - Bell}. Nevertheless, it is applied implicitly
when $\mathcal{S}_{\Theta |\mathbf{x}}=\mathbb{R}$ and explicitly in some
specific examples when $\mathcal{S}_{\Theta |\mathbf{x}}\varsubsetneq
\mathbb{R}$ for which the function $q(\mathbf{x},\theta )$ tends to zero at
the endpoints of $\mathcal{S}_{\Theta |\mathbf{x}}$ (see \cite[Ex. 4.2]%
{BMWZ87}, \cite[Ex. 9]{Van Trees - Bell}).\smallskip \newline
$\bullet $ the following alternative constraint
\begin{equation}
\forall \mathbf{x}\in \mathcal{S}_{\mathcal{X}}:\lim\limits_{\theta
\rightarrow a_{\mathbf{x}}}p\left( \theta |\mathbf{x}\right)
=0=\lim\limits_{\theta \rightarrow a_{\mathbf{x}}}\frac{\partial p\left(
\theta |\mathbf{x}\right) }{\partial \theta }\text{ and }\lim\limits_{\theta
\rightarrow b_{\mathbf{x}}}p\left( \theta |\mathbf{x}\right)
=0=\lim\limits_{\theta \rightarrow b_{\mathbf{x}}}\frac{\partial p\left(
\theta |\mathbf{x}\right) }{\partial \theta },  \label{cond 2}
\end{equation}%
leads to the $\mathrm{BMZB}_{q}$ as well (but not mentioned in \cite[\S 4]%
{BMWZ87}).\smallskip \newline
As conditions (\ref{cond 1}) and (\ref{cond 2}) may hold in many cases,
Proposition 1 highlights the fact that the BMZB is not only a class of BCRBs
(as initially introduced in \cite{BMWZ87}) or weighted BCRBs (so-called in
\cite{Van Trees - Bell}), but rather the general form of tight BCRBs (\ref%
{BMWZB}) when defined as the limiting form of some large-error bounds.

\subsection{Case where $\mathcal{S}_{\Theta |\mathbf{x}}$ is a countable
union of disjoint intervals of $%
\mathbb{R}
$}

Interestingly enough, Proposition 1 and, as a consequence, the $\mathrm{BMZB}%
_{q}$, can be formulated in the general case where $\mathcal{S}_{\Theta |%
\mathbf{x}}$ is a countable union of disjoint intervals $\mathcal{I}_{\Theta
|\mathbf{x}}^{k}$ of $%
\mathbb{R}
$:
\end{subequations}
\begin{equation}
\forall \mathbf{x}\in \mathcal{S}_{\mathcal{X}},~\mathcal{S}_{\Theta |%
\mathbf{x}}=\dbigcup\limits_{k\in \mathcal{K}_{\mathbf{x}}}\mathcal{I}%
_{\Theta |\mathbf{x}}^{k},\text{\quad }\mathcal{I}_{\Theta |\mathbf{x}%
}^{k}\cap \mathcal{I}_{\Theta |\mathbf{x}}^{l}=\varnothing ,\text{~}\forall
k,l\in \mathcal{K}_{\mathbf{x}},k\neq l,  \label{countable union}
\end{equation}%
where $\mathcal{K}_{\mathbf{x}}$ denotes a subset of $%
\mathbb{N}
$.\newline
Indeed, we can state the following:\smallskip \newline
\textbf{Proposition 2} : If $\forall \mathbf{x}\in \mathcal{S}_{\mathcal{X}}$%
:\newline
$\bullet ~\mathcal{S}_{\Theta |\mathbf{x}}$ is a countable union of disjoint
intervals $\mathcal{I}_{\Theta |\mathbf{x}}^{k}$ of $%
\mathbb{R}
$ (\ref{countable union}) with endpoints $a_{\mathbf{x}}^{k},b_{\mathbf{x}%
}^{k}\in \left[ -\infty ,+\infty \right] $, $a_{\mathbf{x}}^{k}<b_{\mathbf{x}%
}^{k}$,\newline
$\bullet ~q\left( \mathbf{x},\theta \right) $ admits a finite limit at
endpoints of $\mathcal{I}_{\Theta |\mathbf{x}}^{k}$,\newline
$\bullet ~g\left( \theta \right) $ is piecewise $\mathcal{C}^{1}$ w.r.t. $%
\theta $ over $\mathcal{I}_{\Theta |\mathbf{x}}^{k}$,\newline
$\bullet ~t\left( \mathbf{x},\theta \right) \triangleq q\left( \mathbf{x}%
,\theta \right) p\left( \theta |\mathbf{x}\right) $ is piecewise $\mathcal{C}%
^{1}$ w.r.t. $\theta $ over $\mathcal{I}_{\Theta |\mathbf{x}}^{k}$ and such
as $\frac{\partial t\left( \mathbf{x},\theta \right) }{\partial \theta }$
admits a finite limit at endpoints of $\mathcal{I}_{\Theta |\mathbf{x}}^{k}$,%
\newline
${\small \bullet }~u\left( \mathbf{x},\theta \right) \triangleq q\left(
\mathbf{x},\theta \right) ^{2}p\left( \theta |\mathbf{x}\right) $ is $%
\mathcal{C}^{2}$ w.r.t. $\theta $ at the vicinity of endpoints of $\mathcal{I%
}_{\Theta |\mathbf{x}}^{k}$ and such as $u\left( \mathbf{x},\theta \right) ,%
\frac{\partial u\left( \mathbf{x},\theta \right) }{\partial \theta }$ and $%
\frac{\partial ^{2}u\left( \mathbf{x},\theta \right) }{\partial ^{2}\theta }$
admit a finite limit at endpoints of $\mathcal{I}_{\Theta |\mathbf{x}}^{k}$,%
\newline
then a necessary and sufficient condition in order to obtain a non trivial $%
\mathrm{BCRB}_{q}$ is:
\begin{subequations}
\begin{equation}
\lim\limits_{\theta \rightarrow a_{\mathbf{x}}^{k}}t\left( \mathbf{x},\theta
\right) =0=\lim\limits_{\theta \rightarrow b_{\mathbf{x}}^{k}}t\left(
\mathbf{x},\theta \right) ,\quad \forall k\in \mathcal{K}_{\mathbf{x}},
\label{cond BCRB non trivial - disconnected subset}
\end{equation}%
leading to:%
\begin{equation}
\mathrm{BCRB}_{q}\left( g\left( \theta \right) \right) =\frac{E_{\mathbf{x}%
,\theta }\left[ \frac{dg\left( \theta \right) }{d\theta }q\left( \mathbf{x}%
,\theta \right) \right] ^{2}}{E_{\mathbf{x},\theta }\left[ \left( \frac{%
\frac{\partial t\left( \mathbf{x},\theta \right) }{\partial \theta }}{%
p\left( \theta |\mathbf{x}\right) }\right) ^{2}\right] +\min \left\{
\begin{array}{l}
\tsum\limits_{k\in \mathcal{K}_{\mathbf{x}}}E_{\mathbf{x}}\left[ \frac{5}{2}%
\lim\limits_{\theta \rightarrow a_{\mathbf{x}}^{k}}v\left( \mathbf{x},\theta
\right) -\frac{1}{2}\lim\limits_{\theta \rightarrow b_{\mathbf{x}%
}^{k}}v\left( \mathbf{x},\theta \right) \right] \smallskip , \\
\tsum\limits_{k\in \mathcal{K}_{\mathbf{x}}}E_{\mathbf{x}}\left[ \frac{1}{2}%
\lim\limits_{\theta \rightarrow a_{\mathbf{x}}^{k}}v\left( \mathbf{x},\theta
\right) -\frac{5}{2}\lim\limits_{\theta \rightarrow b_{\mathbf{x}%
}^{k}}v\left( \mathbf{x},\theta \right) \right]
\end{array}%
\right\} }  \label{BCRB - disconnected subset}
\end{equation}%
where $v\left( \mathbf{x},\theta \right) =q\left( \mathbf{x},\theta \right)
^{2}\frac{\partial p\left( \theta |\mathbf{x}\right) }{\partial \theta }$%
.\medskip \newline
\textit{Proof:} see Appendix \ref{A: Case of a countable union of disjoint
intervals}.\medskip \newline
Choosing $q\left( \mathbf{x},\theta \right) $ such that
\end{subequations}
\begin{subequations}
\begin{equation}
\forall \mathbf{x}\in \mathcal{S}_{\mathcal{X}},\forall k\in \mathcal{K}_{%
\mathbf{x}}:\lim\limits_{\theta \rightarrow a_{\mathbf{x}}^{k}}q\left(
\mathbf{x},\theta \right) =\lim\limits_{\theta \rightarrow b_{\mathbf{x}%
}^{k}}q\left( \mathbf{x},\theta \right) =0,  \label{cond 1 - bis}
\end{equation}%
then (\ref{BCRB - disconnected subset}) reduces to $\mathrm{BMZB}_{q}\left(
g\left( \theta \right) \right) $ (\ref{BMWZB}). Last, note that the
following alternative constraints
\begin{equation}
\lim\limits_{\theta \rightarrow a_{\mathbf{x}}^{k}}p\left( \theta |\mathbf{x}%
\right) =\lim\limits_{\theta \rightarrow a_{\mathbf{x}}^{k}}\frac{\partial
p\left( \theta |\mathbf{x}\right) }{\partial \theta }=0\text{ and }%
\lim\limits_{\theta \rightarrow b_{\mathbf{x}}^{k}}p\left( \theta |\mathbf{x}%
\right) =\lim\limits_{\theta \rightarrow b_{\mathbf{x}}^{k}}\frac{\partial
p\left( \theta |\mathbf{x}\right) }{\partial \theta }=0,\quad \forall k\in
\mathcal{K}_{\mathbf{x}},  \label{cond 2 - bis}
\end{equation}%
leads to the $\mathrm{BMZB}_{q}$ (\ref{BMWZB}) as well.

\section{Examples of Bayesian lower bounds of the Proposed class}

\subsection{Reformulation of existing Bayesian bounds}

We show in this section that expression (\ref{BLB}), with a judicious choice
of the function $q$, allows for a general formulation of existing BLBs
whatever $\mathcal{S}_{\Theta |\mathbf{x}}\subset
\mathbb{R}
$, including naturally the cases of a bounded connected subset of $%
\mathbb{R}
$ (see Section \ref{S: example}) or a disjoint subset of $%
\mathbb{R}
$ \cite{Ben-Haim - Eldar}.\medskip

\subsubsection{Case of the Weiss-Weinstein lower bound}

\quad \newline
In order to obtain the WWB, we specify, for $s\in ]0,1[$, the function:
\end{subequations}
\begin{subequations}
\begin{equation}
q_{\mathrm{WW}}^{h,s}\left( \mathbf{x},\theta \right) =%
\begin{cases}
\left( \frac{p\left( \theta -h|\mathbf{x}\right) }{p\left( \theta |\mathbf{x}%
\right) }\right) ^{1-s}1_{\mathcal{S}_{\Theta |\mathbf{x}}}\left( \theta
-h\right) 1_{\mathcal{S}_{\Theta |\mathbf{x}}}\left( \theta \right) , &
\text{if }\left( \mathbf{x},\theta \right) \in \mathcal{S}_{\mathcal{X}%
,\Theta } \\
0, & \text{otherwise}%
\end{cases}%
.  \label{WWB - q}
\end{equation}%
Consequently, using $q_{\mathrm{WW}}^{h,s}\left( \mathbf{x},\theta \right) $
into (\ref{psi(h,f) - WWF}), one obtains the function:%
\begin{equation}
\psi _{\mathrm{WW}}^{h,s}\left( \mathbf{x},\theta \right) =%
\begin{cases}
\left( \left( \frac{p\left( \theta +h|\mathbf{x}\right) }{p\left( \theta |%
\mathbf{x}\right) }\right) ^{s}1_{\mathcal{S}_{\Theta |\mathbf{x}}}\left(
\theta +h\right) -\left( \frac{p\left( \theta -h|\mathbf{x}\right) }{p\left(
\theta |\mathbf{x}\right) }\right) ^{1-s}1_{\mathcal{S}_{\Theta |\mathbf{x}%
}}\left( \theta -h\right) \right) 1_{\mathcal{S}_{\Theta |\mathbf{x}}}\left(
\theta \right) , & \text{if }\left( \mathbf{x},\theta \right) \in \mathcal{S}%
_{\mathcal{X},\Theta } \\
0, & \text{otherwise}%
\end{cases}%
,  \label{WWB - psi}
\end{equation}%
and an explicit form of WWB introduced in \cite{Weiss - Weinstein 85} is:
\end{subequations}
\begin{subequations}
\begin{gather}
\mathrm{WWB}\left( g\left( \theta \right) \right) =\sup_{s\in ]0,1[,h\in
\mathbb{R}:~\psi _{\mathrm{WW}}^{h,s}\left( .\right) \in \mathcal{L}_{2}(%
\mathcal{S}_{\mathcal{X},\Theta })}\left\{ \mathrm{WWB}^{h,s}\left( g\left(
\theta \right) \right) \right\} ,  \label{WWB} \\
\mathrm{WWB}^{h,s}\left( g\left( \theta \right) \right) =\frac{E_{\mathbf{x}%
,\theta }\left[ \left( g\left( \theta -h\right) -g\left( \theta \right)
\right) \left( \frac{p\left( \theta -h|\mathbf{x}\right) }{p\left( \theta |%
\mathbf{x}\right) }\right) ^{1-s}1_{\mathcal{S}_{\Theta |\mathbf{x}}}\left(
\theta -h\right) 1_{\mathcal{S}_{\Theta |\mathbf{x}}}\left( \theta \right) %
\right] ^{2}}{E_{\mathbf{x},\theta }\left[ \left( \left( \frac{p\left(
\theta +h|\mathbf{x}\right) }{p\left( \theta |\mathbf{x}\right) }\right)
^{s}1_{\mathcal{S}_{\Theta |\mathbf{x}}}\left( \theta +h\right) -\left(
\frac{p\left( \theta -h|\mathbf{x}\right) }{p\left( \theta |\mathbf{x}%
\right) }\right) ^{1-s}1_{\mathcal{S}_{\Theta |\mathbf{x}}}\left( \theta
-h\right) \right) ^{2}1_{\mathcal{S}_{\Theta |\mathbf{x}}}\left( \theta
\right) \right] }  \label{WWB(h,s) explicit form}
\end{gather}%
It is worth noting that the use of the compact form \cite[(20-21)]{Weiss -
Weinstein 88} can be a source of error in the formulation of the integration
domains involved in the computations of the various expectations when $%
\mathcal{S}_{\Theta |\mathbf{x}}$ is a bounded connected subset of $%
\mathbb{R}
$ or a disjoint subset of $%
\mathbb{R}
$, as exemplified in \cite{Ben-Haim - Eldar}.\bigskip

\subsubsection{Case of the Bobrovsky-Zakai bound}

\quad \newline
In order to obtain the BZB, we set $q_{\mathrm{BZ}}^{h}\left( \mathbf{x}%
,\theta \right) =\frac{1}{h}$ leading to:
\end{subequations}
\begin{equation}
\psi _{\mathrm{BZ}}^{h}\left( \mathbf{x},\theta \right) =%
\begin{cases}
\frac{1}{h}\left( \frac{p\left( \theta +h|\mathbf{x}\right) }{p\left( \theta
|\mathbf{x}\right) }1_{\mathcal{S}_{\Theta |\mathbf{x}}}\left( \theta
+h\right) -1_{\mathcal{S}_{\Theta |\mathbf{x}}}\left( \theta -h\right)
\right) 1_{\mathcal{S}_{\Theta |\mathbf{x}}}\left( \theta \right) , & \text{%
if }\left( \mathbf{x},\theta \right) \in \mathcal{S}_{\mathcal{X},\Theta }
\\
0, & \text{otherwise}%
\end{cases}%
.  \label{BZB - psi}
\end{equation}%
Consequently, a regularized explicit form of BZB is given by:
\begin{subequations}
\begin{gather}
\mathrm{BZB}\left( g\left( \theta \right) \right) =\sup_{h\in \mathbb{R}%
:~\psi _{\mathrm{BZ}}^{h}\left( .\right) \in \mathcal{L}_{2}(\mathcal{S}_{%
\mathcal{X},\Theta })}\left\{ \mathrm{BZB}^{h}\left( g\left( \theta \right)
\right) \right\} ,  \label{BZB} \\
\mathrm{BZB}^{h}\left( g\left( \theta \right) \right) =\frac{E_{\mathbf{x}%
,\theta }\left[ \left( \frac{g\left( \theta -h\right) -g\left( \theta
\right) }{h}\right) 1_{\mathcal{S}_{\Theta |\mathbf{x}}}\left( \theta
-h\right) 1_{\mathcal{S}_{\Theta |\mathbf{x}}}\left( \theta \right) \right]
^{2}}{E_{\mathbf{x},\theta }\left[ \left( \frac{p\left( \theta +h|\mathbf{x}%
\right) 1_{\mathcal{S}_{\Theta |\mathbf{x}}}\left( \theta +h\right) -p\left(
\theta |\mathbf{x}\right) 1_{\mathcal{S}_{\Theta |\mathbf{x}}}\left( \theta
-h\right) }{hp\left( \theta |\mathbf{x}\right) }\right) ^{2}1_{\mathcal{S}%
_{\Theta |\mathbf{x}}}\left( \theta \right) \right] },  \label{BZB - h}
\end{gather}%
which is a generalization of the bound introduced in \cite{Bobrovsky - Zakai}
whatever $\mathcal{S}_{\Theta |\mathbf{x}}$. From a practical viewpoint, it
is a noticeable result, since the BZB is the easiest to use "large-error"
bound, but was believed to be inapplicable where $\mathcal{S}_{\Theta |%
\mathbf{x}}$ is a bounded connected subset of $%
\mathbb{R}
$ \cite[Section II]{Bobrovsky - Zakai}\cite[p682]{Weiss - Weinstein 85}\cite[%
p340]{Weiss - Weinstein 88}\cite[p39]{Van Trees - Bell}. Moreover, since, $%
\forall y>0,\underset{s\rightarrow 1^{-}}{\lim }y^{1-s}=1$, therefore, $%
\forall h\in
\mathbb{R}
$ and $\forall \left( \mathbf{x},\theta \right) \in \mathcal{S}_{\mathcal{X}%
,\Theta }$:
\end{subequations}
\begin{subequations}
\begin{equation}
\underset{s\rightarrow 1^{-}}{\lim }q_{\mathrm{WW}}^{h,s}\left( \mathbf{x}%
,\theta \right) =1_{\mathcal{S}_{\Theta |\mathbf{x}}}\left( \theta -h\right)
1_{\mathcal{S}_{\Theta |\mathbf{x}}}\left( \theta \right) ,
\label{limit of q WWB for s = 1}
\end{equation}%
leading to:
\begin{equation}
\underset{s\rightarrow 1^{-}}{\lim }\psi _{\mathrm{WW}}^{h,s}\left( \mathbf{x%
},\theta \right) =\left( \frac{p\left( \theta +h|\mathbf{x}\right) }{p\left(
\theta |\mathbf{x}\right) }1_{\mathcal{S}_{\Theta |\mathbf{x}}}\left( \theta
+h\right) -1_{\mathcal{S}_{\Theta |\mathbf{x}}}\left( \theta -h\right)
\right) 1_{\mathcal{S}_{\Theta |\mathbf{x}}}\left( \theta \right) =h\psi _{%
\mathrm{BZ}}^{h}\left( \mathbf{x},\theta \right) ,
\end{equation}%
and:
\begin{equation}
\underset{s\rightarrow 1^{-}}{\lim }\mathrm{WWB}^{h,s}\left( g\left( \theta
\right) \right) =\mathrm{BZB}^{h}\left( g\left( \theta \right) \right) ,
\label{convergence of WWB towards BZB}
\end{equation}%
which is an extension of the result introduced in \cite{Weiss - Weinstein 88}
whatever $\mathcal{S}_{\Theta |\mathbf{x}}$.\bigskip

\subsubsection{Generalization}

\quad \newline
It is straightforward to extend the derivation of all the other existing
BLBs mentioned in \cite{RFLRN08} and \cite{Todros - Tabrikian - IT -
Bayesian} whatever $\mathcal{S}_{\Theta |\mathbf{x}}$, namely the historical
BCRB, the BMZB, the Bayesian Bhattacharayya bound \cite{Weiss - Weinstein 88}%
, the Reuven-Messer bound \cite{Reuven - Messer}, the combined Cram\'{e}%
r-Rao/Weiss-Weinstein bound \cite{Bell - Van Trees}, the Bayesian Abel bound
\cite{Renaux AB}, and the Bayesian Todros-Tabrikian bound \cite{Todros -
Tabrikian - IT - Bayesian}, by updating the definitions of $\nu _{RM}\left(
\mathbf{x},\theta ,\tau \right) $ \cite[(32)]{Todros - Tabrikian - IT -
Bayesian} and $\nu _{WW}\left( \mathbf{x},\theta ,\tau \right) $ \cite[(33)]%
{Todros - Tabrikian - IT - Bayesian} as follows:
\end{subequations}
\begin{equation}
\nu _{RM}\left( \mathbf{x},\theta ,\tau \right) =\psi _{\mathrm{BZ}}^{\tau
}\left( \mathbf{x},\theta \right) ,\quad \nu _{WW}\left( \mathbf{x},\theta
,\tau \right) =\psi _{\mathrm{WW}}^{\tau ,\beta \left( \tau \right) }\left(
\mathbf{x},\theta \right) .
\end{equation}

\subsection{Modified Weiss-Weinstein and Bobrovsky-Zakai lower bounds\label%
{S: A modified Weiss-Weinstein lower bound}}

It is now known and exemplified \cite{BMWZ87}\cite[p36]{Van Trees - Bell}
that the BMZB not only allows to derive a non trivial BCRB in cases where
the historical BCRB is trivial but may also provides a tighter bound than
the historical BCRB in the asymptotic region. Since the limiting form of the
WWB (\ref{WWB}-\ref{WWB(h,s) explicit form}) and BZB (\ref{BZB}-\ref{BZB - h}%
) is the historical BCRB, it would seem sensible to define modified WWB and
BZB which limiting form is the BMZB, in expectation of an increased
tightness in the threshold region as well. In that perspective, a modified
WWB, denoted $\mathrm{WWB}_{q}$ in the following, which limiting form is $%
\mathrm{BMZB}_{q}$ (\ref{BMWZB}), can be obtained by modifying the
definition of $q_{\mathrm{WW}}^{h,s}\left( \mathbf{x},\theta \right) $ (\ref%
{WWB - q}) as follows:
\begin{equation}
q_{\mathrm{MWW}}^{h,s}\left( \mathbf{x},\theta \right) =%
\begin{cases}
\left( \frac{p\left( \theta -h|\mathbf{x}\right) }{p\left( \theta |\mathbf{x}%
\right) }\right) ^{1-s}q\left( \mathbf{x},\theta \right) 1_{\mathcal{S}%
_{\Theta |\mathbf{x}}}\left( \theta -h\right) 1_{\mathcal{S}_{\Theta |%
\mathbf{x}}}\left( \theta \right) , & \text{if }\left( \mathbf{x},\theta
\right) \in \mathcal{S}_{\mathcal{X},\Theta } \\
0, & \text{otherwise}%
\end{cases}%
,
\end{equation}%
provided that one of the conditions (\ref{cond 1}), (\ref{cond 2}), (\ref%
{cond 1 - bis}), (\ref{cond 2 - bis}) holds, since, according to (\ref{limit
of q WWB for s = 1}):%
\begin{equation}
\underset{s\rightarrow 1^{-}}{\lim }q_{\mathrm{MWW}}^{h,s}\left( \mathbf{x}%
,\theta \right) =q\left( \mathbf{x},\theta \right) 1_{\mathcal{S}_{\Theta |%
\mathbf{x}}}\left( \theta -h\right) 1_{\mathcal{S}_{\Theta |\mathbf{x}%
}}\left( \theta \right) .
\end{equation}%
Thus:
\begin{subequations}
\begin{gather}
\mathrm{WWB}_{q}\left( g\left( \theta \right) \right) =\sup_{s\in ]0,1[,h\in
\mathbb{R}:~\psi _{\mathrm{MWW}}^{h,s}\left( .\right) \in \mathcal{L}_{2}(%
\mathcal{S}_{\mathcal{X},\Theta })}\left\{ \mathrm{WWB}_{q}^{h,s}\left(
g\left( \theta \right) \right) \right\} ,  \label{MOD WWB - Gen} \\
\mathrm{WWB}_{q}^{h,s}\left( g\left( \theta \right) \right) =\frac{E_{%
\mathbf{x},\theta }\left[ \left( g\left( \theta -h\right) -g\left( \theta
\right) \right) \left( \frac{p\left( \theta -h|\mathbf{x}\right) }{p\left(
\theta |\mathbf{x}\right) }\right) ^{1-s}q\left( \mathbf{x},\theta \right)
1_{\mathcal{S}_{\Theta |\mathbf{x}}}\left( \theta -h\right) 1_{\mathcal{S}%
_{\Theta |\mathbf{x}}}\left( \theta \right) \right] ^{2}}{E_{\mathbf{x}%
,\theta }\left[ \left(
\begin{array}{l}
\left( \frac{p\left( \theta +h|\mathbf{x}\right) }{p\left( \theta |\mathbf{x}%
\right) }\right) ^{s}q\left( \mathbf{x},\theta +h\right) 1_{\mathcal{S}%
_{\Theta |\mathbf{x}}}\left( \theta +h\right) - \\
\left( \frac{p\left( \theta -h|\mathbf{x}\right) }{p\left( \theta |\mathbf{x}%
\right) }\right) ^{1-s}q\left( \mathbf{x},\theta \right) 1_{\mathcal{S}%
_{\Theta |\mathbf{x}}}\left( \theta -h\right)
\end{array}%
\right) ^{2}1_{\mathcal{S}_{\Theta |\mathbf{x}}}\left( \theta \right) \right]
}  \label{MOD WWB(h,s,q) - Gen}
\end{gather}%
Note that the usual WWB is obtained for $q\left( \mathbf{x},\theta \right)
=1_{\mathcal{S}_{\mathcal{X},\Theta }}\left( \mathbf{x},\theta \right) $ and
the modified BZB, denoted $\mathrm{BZB}_{q}$ in the following, is obtained
for $q_{\mathrm{MBZ}}^{h}\left( \mathbf{x},\theta \right) =\underset{%
s\rightarrow 1^{-}}{\lim }q_{\mathrm{MWW}}^{h,s}\left( \mathbf{x},\theta
\right) $, leading to:
\end{subequations}
\begin{subequations}
\begin{gather}
\mathrm{BZB}_{q}\left( g\left( \theta \right) \right) =\sup_{h\in \mathbb{R}%
:~\psi _{\mathrm{MBZ}}^{h}\left( .\right) \in \mathcal{L}_{2}(\mathcal{S}_{%
\mathcal{X},\Theta })}\left\{ \mathrm{BZB}_{q}^{h}\left( g\left( \theta
\right) \right) \right\} ,  \label{MOD BZB} \\
\mathrm{BZB}_{q}^{h}\left( g\left( \theta \right) \right) =\frac{E_{\mathbf{x%
},\theta }\left[ \left( \frac{g\left( \theta -h\right) -g\left( \theta
\right) }{h}\right) 1_{\mathcal{S}_{\Theta |\mathbf{x}}}\left( \theta
-h\right) 1_{\mathcal{S}_{\Theta |\mathbf{x}}}\left( \theta \right) \right]
^{2}}{E_{\mathbf{x},\theta }\left[ \left( \frac{p\left( \theta +h|\mathbf{x}%
\right) q\left( \mathbf{x},\theta +h\right) 1_{\mathcal{S}_{\Theta |\mathbf{x%
}}}\left( \theta +h\right) -p\left( \theta |\mathbf{x}\right) q\left(
\mathbf{x},\theta \right) 1_{\mathcal{S}_{\Theta |\mathbf{x}}}\left( \theta
-h\right) }{hp\left( \theta |\mathbf{x}\right) }\right) ^{2}1_{\mathcal{S}%
_{\Theta |\mathbf{x}}}\left( \theta \right) \right] }.  \label{MOD BZB(h)}
\end{gather}

\section{Application to the Gaussian observation model with parameterized
mean and uniform prior\label{S: example}}

This section is dedicated to exemplify some of the results introduced above
with a reference problem in signal processing: the Gaussian observation
model with a parameterized mean depending on a random parameter with uniform
prior. For numerical evaluations, we focus on the estimation of a single
tone. Thus the parametric model under consideration is:
\end{subequations}
\begin{equation}
\mathbf{x}=\left( x_{1},\ldots ,x_{N}\right) ^{T}=\boldsymbol{m}\left(
\theta \right) +\mathbf{n},\quad p\left( \mathbf{x}|\theta \right) =\frac{%
e^{-\frac{\left\Vert \mathbf{x}-\boldsymbol{m}\left( \theta \right)
\right\Vert ^{2}}{\sigma _{\mathbf{n}}^{2}}}}{\left( \pi \sigma _{\mathbf{n}%
}^{2}\right) ^{N}},\quad p\left( \theta \right) =\frac{1_{\Theta }\left(
\theta \right) }{b-a},  \label{obs parameterized mean}
\end{equation}%
where $\mathcal{S}_{\mathcal{X}|\theta }=\mathcal{S}_{\mathcal{X}}=\mathbb{C}%
^{N}$ and $\mathcal{S}_{\Theta |\mathbf{x}}=\mathcal{S}_{\Theta }=\left[ a,b%
\right] $. \newline
In the case of single tone estimation, $\boldsymbol{m}\left( \theta \right)
=\alpha \left( 1,e^{j2\pi \theta },\dots ,e^{j(N-1)2\pi \theta }\right) ^{T}$%
, $\alpha \in
\mathbb{C}
$, $\left[ a,b\right] \triangleq \left[ 0,1\right] $, and $g\left( \theta
\right) \triangleq \theta $.\newline
A motivation for choosing the parametric model (\ref{obs parameterized mean}%
) is the belief in the open literature that both the BCRB and the BZB are
inapplicable in that case \cite[Section II]{Bobrovsky - Zakai}\cite[p682]%
{Weiss - Weinstein 85}\cite[p340]{Weiss - Weinstein 88}\cite[p39]{Van Trees
- Bell}.

\subsection{The WWB and its limiting form}

For the parametric model (\ref{obs parameterized mean}), the WWB (\ref{WWB}-%
\ref{WWB(h,s) explicit form}) is given by \cite[Section 4]{Vu et al}:
\begin{subequations}
\begin{gather}
\mathrm{WWB}\left( g\left( \theta \right) \right) =\underset{s\in
]0,1[,|h|<b-a}{\sup }\left\{ \mathrm{WWB}^{h,s}\left( g\left( \theta \right)
\right) \right\} ,  \label{WWB ST 1-0} \\
\mathrm{WWB}^{h,s}\left( g\left( \theta \right) \right) =\frac{E_{\theta }%
\left[ \left( g\left( \theta -h\right) -g\left( \theta \right) \right) e^{-%
\frac{\left( 1-s\right) s}{\sigma _{\mathbf{n}}^{2}}\left\Vert \mathbf{m}%
\left( \theta -h\right) -\boldsymbol{m}\left( \theta \right) \right\Vert
^{2}}1_{\mathcal{S}_{\Theta }}\left( \theta -h\right) \right] ^{2}}{\left(
\begin{array}{l}
~E_{\theta }\left[ e^{-\frac{2s\left( 1-2s\right) }{\sigma _{\mathbf{n}}^{2}}%
\left\Vert \boldsymbol{m}\left( \theta +h\right) -\boldsymbol{m}\left(
\theta \right) \right\Vert ^{2}}1_{\mathcal{S}_{\Theta }}\left( \theta
+h\right) \right] +\smallskip  \\
~E_{\theta }\left[ e^{-\frac{2\left( 1-s\right) \left( 2s-1\right) }{\sigma
_{\mathbf{n}}^{2}}\left\Vert \boldsymbol{m}\left( \theta -h\right) -%
\boldsymbol{m}\left( \theta \right) \right\Vert ^{2}}1_{\mathcal{S}_{\Theta
}}\left( \theta -h\right) \right] -\smallskip  \\
2E_{\theta }\left[ e^{-\frac{s\left( 1-s\right) }{\sigma _{\mathbf{n}}^{2}}%
\left\Vert \boldsymbol{m}\left( \theta +h\right) -\boldsymbol{m}\left(
\theta -h\right) \right\Vert ^{2}}1_{\mathcal{S}_{\Theta }}\left( \theta
-h\right) 1_{\mathcal{S}_{\Theta }}\left( \theta +h\right) \right]
\end{array}%
\right) }.  \label{WWB ST 1-1}
\end{gather}%
As stated by \textit{Proposition 1}, since $t\left( \mathbf{x},\theta
\right) \triangleq q\left( \mathbf{x},\theta \right) p\left( \theta |\mathbf{%
x}\right) =1_{\mathcal{S}_{\mathcal{X},\Theta }}\left( \mathbf{x},\theta
\right) 1_{\Theta }\left( \theta \right) =1_{\Theta }\left( \theta \right) $%
, thus $t\left( \mathbf{x},\theta \right) $ does not verify (\ref{cond BCRB
non trivial - connected subset}) and the associated generalized $\mathrm{BCRB%
}_{1_{\mathcal{S}_{\mathcal{X},\Theta }}}$ (\ref{GBCRB}) is trivial.\newline
Indeed, since $\forall h:|h|<1$, $E_{\theta }\left[ 1_{\mathcal{S}_{\Theta
}}\left( \theta \pm h\right) \right] =1-\left\vert h\right\vert $, $%
E_{\theta }\left[ 1_{\mathcal{S}_{\Theta }}\left( \theta +h\right) 1_{%
\mathcal{S}_{\Theta }}\left( \theta -h\right) \right] =\sup \left\{
1-2\left\vert h\right\vert ,0\right\} $, and $\underset{h_{1},h_{2}%
\rightarrow 0}{\lim }\left\Vert \boldsymbol{m}\left( \theta +h_{1}\right) -%
\boldsymbol{m}\left( \theta +h_{2}\right) \right\Vert ^{2}=\left\Vert \frac{%
\partial \mathbf{m}\left( \theta \right) }{\partial \theta }\right\Vert
^{2}(h_{1}-h_{2})^{2}$, then:
\end{subequations}
\begin{subequations}
\begin{equation}
\underset{h\rightarrow 0}{\lim }\mathrm{WWB}^{h,s}\left( g\left( \theta
\right) \right) =\frac{h^{2}E_{\theta }\left[ \frac{\partial g\left( \theta
\right) }{\partial \theta }\right] ^{2}}{2\left\vert h\right\vert
+2h^{2}E_{\theta }\left[ \frac{\left\Vert \frac{\partial \mathbf{m}\left(
\theta \right) }{\partial \theta }\right\Vert ^{2}}{\sigma _{\mathbf{n}}^{2}}%
\right] }=\frac{E_{\theta }\left[ \frac{\partial g\left( \theta \right) }{%
\partial \theta }\right] ^{2}}{\frac{2}{\left\vert h\right\vert }+2E_{\theta
}\left[ \frac{\left\Vert \frac{\partial \mathbf{m}\left( \theta \right) }{%
\partial \theta }\right\Vert ^{2}}{\sigma _{\mathbf{n}}^{2}}\right] },
\label{limit WWB(h,s)}
\end{equation}%
and (\ref{GBCRB})(\ref{convergence of WWB towards BZB}):
\begin{equation}
\mathrm{BCRB}_{1_{\mathcal{S}_{\mathcal{X},\Theta }}}\left( g\left( \theta
\right) \right) =\underset{h\rightarrow 0}{\lim }\mathrm{BZB}^{h}\left(
g\left( \theta \right) \right) =\underset{h\rightarrow 0}{\lim }\left(
\underset{s\rightarrow 1^{-}}{\lim }\mathrm{WWB}^{h,s}\left( g\left( \theta
\right) \right) \right) =\underset{s\rightarrow 1^{-}}{\lim }\left( \underset%
{h\rightarrow 0}{\lim }\mathrm{WWB}^{h,s}\left( g\left( \theta \right)
\right) \right) =0.  \label{WWB limiting form}
\end{equation}%
\medskip

\subsection{Some BMZBs and their associated modified WWBs}

We consider the family of $\mathrm{BMZB}_{q}\left( g\left( \theta \right)
\right) $ (\ref{BMWZB}) obtained where $q\left( \mathbf{x},\theta \right)
\triangleq q\left( \theta \right) $ satisfying (\ref{cond 1}):
\end{subequations}
\begin{equation}
\mathrm{BMZB}_{q}\left( g\left( \theta \right) \right) =\frac{E_{\theta }%
\left[ \frac{dg\left( \theta \right) }{d\theta }q\left( \theta \right) %
\right] ^{2}}{E_{\theta }\left[ E_{\mathbf{x}|\theta }\left[ \left( \frac{%
\partial q\left( \theta \right) }{\partial \theta }+q\left( \theta \right)
\frac{\partial \ln p\left( \theta |\mathbf{x}\right) }{\partial \theta }%
\right) ^{2}\right] \right] },\quad \lim\limits_{\theta \rightarrow
a}q\left( \theta \right) =0=\lim\limits_{\theta \rightarrow b}q\left( \theta
\right) .
\end{equation}%
Then, on one hand:
\begin{subequations}
\begin{equation}
E_{\mathbf{x}|\theta }\left[ \left( \frac{\partial q\left( \theta \right) }{%
\partial \theta }+q\left( \theta \right) \frac{\partial \ln p\left( \theta |%
\mathbf{x}\right) }{\partial \theta }\right) ^{2}\right] =E_{\mathbf{x}%
|\theta }\left[ \left( \frac{\partial q\left( \theta \right) }{\partial
\theta }\right) ^{2}+q\left( \theta \right) ^{2}\left( \frac{\partial \ln
p\left( \theta |\mathbf{x}\right) }{\partial \theta }\right) ^{2}+2\frac{%
\partial q\left( \theta \right) }{\partial \theta }q\left( \theta \right)
\frac{\partial \ln p\left( \theta |\mathbf{x}\right) }{\partial \theta }%
\right]
\end{equation}%
and, on the other hand, $\forall \theta \in \mathcal{S}_{\Theta |\mathbf{x}}$
:%
\begin{equation}
E_{\mathbf{x}|\theta }\left[ \frac{\partial \ln p\left( \theta |\mathbf{x}%
\right) }{\partial \theta }\right] =\frac{\partial \ln p\left( \theta
\right) }{\partial \theta }=0,~E_{\mathbf{x}|\theta }\left[ \left( \frac{%
\partial \ln p\left( \theta |\mathbf{x}\right) }{\partial \theta }\right)
^{2}\right] =-E_{\mathbf{x}|\theta }\left[ \frac{\partial ^{2}\ln p\left(
\mathbf{x}|\theta \right) }{\partial \theta ^{2}}\right] =\frac{2}{\sigma _{%
\mathbf{n}}^{2}}\left\Vert \frac{\partial \boldsymbol{m}\left( \theta
\right) }{\partial \theta }\right\Vert ^{2}.
\end{equation}%
Consequently,
\begin{equation}
E_{\mathbf{x}|\theta }\left[ \left( \frac{\partial q\left( \theta \right) }{%
\partial \theta }+q\left( \theta \right) \frac{\partial \ln p\left( \theta |%
\mathbf{x}\right) }{\partial \theta }\right) ^{2}\right] =\left( \frac{%
\partial q\left( \theta \right) }{\partial \theta }\right) ^{2}+\frac{%
2q\left( \theta \right) ^{2}}{\sigma _{\mathbf{n}}^{2}}\left\Vert \frac{%
\partial \boldsymbol{m}\left( \theta \right) }{\partial \theta }\right\Vert
^{2},
\end{equation}%
and a tighter $\mathrm{BCRB}\left( \theta \right) $ related to the
parametric model (\ref{obs parameterized mean}) can be defined as:
\end{subequations}
\begin{subequations}
\begin{gather}
\mathrm{BCRB}\left( \theta \right) =\underset{q_{l}\left( .\right) \text{
s.t. (\ref{cond 1})},1\leq l\leq L}{\sup }\left\{ \mathrm{BMZB}%
_{q_{l}}\left( \theta \right) \right\} ,  \label{BCRB ST} \\
\mathrm{BMZB}_{q_{l}}\left( g\left( \theta \right) \right) =\frac{E_{\theta }%
\left[ \frac{dg\left( \theta \right) }{d\theta }q_{l}\left( \theta \right) %
\right] ^{2}}{E_{\theta }\left[ \left( \frac{\partial q_{l}\left( \theta
\right) }{\partial \theta }\right) ^{2}\right] +E_{\theta }\left[ c\left(
\theta \right) q_{l}\left( \theta \right) ^{2}\right] },\quad c\left( \theta
\right) =\frac{2}{\sigma _{\mathbf{n}}^{2}}\left\Vert \frac{\partial
\boldsymbol{m}\left( \theta \right) }{\partial \theta }\right\Vert ^{2}.
\label{BMWZBq ST}
\end{gather}%
Furthermore, the associated modified WWB (\ref{MOD WWB - Gen}-\ref{MOD
WWB(h,s,q) - Gen}) becomes:
\end{subequations}
\begin{subequations}
\begin{equation}
\mathrm{WWB}_{q}\left( \theta \right) =\underset{s\in ]0,1[,|h|<b-a}{\sup }%
\left\{ \mathrm{WWB}_{q}^{h,s}\left( \theta \right) \right\} ,
\label{MOD WWB}
\end{equation}%
\begin{equation}
\mathrm{WWB}_{q}^{h,s}\left( \theta \right) =\frac{E_{\theta }\left[ \left(
g\left( \theta -h\right) -g\left( \theta \right) \right) e^{-\frac{\left(
1-s\right) s}{\sigma _{\mathbf{n}}^{2}}\left\Vert \mathbf{m}\left( \theta
-h\right) -\boldsymbol{m}\left( \theta \right) \right\Vert ^{2}}q\left(
\theta \right) 1_{\mathcal{S}_{\Theta }}\left( \theta -h\right) \right] ^{2}%
}{\left(
\begin{array}{l}
~E_{\theta }\left[ e^{-\frac{2s\left( 1-2s\right) }{\sigma _{\mathbf{n}}^{2}}%
\left\Vert \boldsymbol{m}\left( \theta +h\right) -\boldsymbol{m}\left(
\theta \right) \right\Vert ^{2}}q\left( \theta +h\right) ^{2}1_{\mathcal{S}%
_{\Theta }}\left( \theta +h\right) \right] \smallskip + \\
~E_{\theta }\left[ e^{-\frac{2\left( 1-s\right) \left( 2s-1\right) }{\sigma
_{\mathbf{n}}^{2}}\left\Vert \boldsymbol{m}\left( \theta -h\right) -%
\boldsymbol{m}\left( \theta \right) \right\Vert ^{2}}q\left( \theta \right)
^{2}1_{\mathcal{S}_{\Theta }}\left( \theta -h\right) \right] \smallskip - \\
2E_{\theta }\left[ e^{-\frac{s\left( 1-s\right) }{\sigma _{\mathbf{n}}^{2}}%
\left\Vert \boldsymbol{m}\left( \theta +h\right) -\boldsymbol{m}\left(
\theta -h\right) \right\Vert ^{2}}q\left( \theta +h\right) q\left( \theta
\right) 1_{\mathcal{S}_{\Theta }}\left( \theta -h\right) 1_{\mathcal{S}%
_{\Theta }}\left( \theta +h\right) \right]
\end{array}%
\right) }.  \label{MOD WWB(h,s,q)}
\end{equation}%
As an example, two possible choices of the function $q\left( \theta \right) $
are:
\end{subequations}
\begin{subequations}
\begin{equation}
q_{1}^{\delta }\left( \theta \right) =\left\{
\begin{array}{ll}
\frac{1}{2}\left( 1+\sin \left( \pi \left( \frac{\theta }{\delta }-\frac{1}{2%
}\right) \right) \right) , & \text{if }\theta \in \left[ 0,\delta \right]
\\
1, & \text{if }\theta \in \left] \delta ,1-\delta \right[  \\
\frac{1}{2}\left( 1-\sin \left( \pi \left( \frac{\theta -1+\delta }{\delta }-%
\frac{1}{2}\right) \right) \right) , & \text{if }\theta \in \left[ 1-\delta
,1\right]  \\
0, & \text{otherwise}%
\end{array}%
\right. ,  \label{q1}
\end{equation}%
and \cite[p36]{Van Trees - Bell}\cite[p1433]{BMWZ87}:%
\begin{equation}
q_{2}^{\alpha }\left( \theta \right) =%
\begin{cases}
\theta ^{\alpha -1}\left( 1-\theta \right) ^{\alpha -1}, & \text{if }\theta
\in \left[ 0,1\right]  \\
0, & \text{otherwise }%
\end{cases}%
,\quad \alpha >\frac{3}{2}.\qquad \qquad .  \label{q2}
\end{equation}%
In the case of single tone estimation, (\ref{BMWZBq ST}) reduces to (after a
few lines of calculus):
\end{subequations}
\begin{subequations}
\begin{eqnarray}
\mathrm{BMZB}_{q_{1}^{\delta }}\left( \theta \right)  &=&\frac{\left(
1-\delta \right) ^{2}}{\frac{\pi ^{2}}{4\delta }+\frac{4}{3}\pi ^{2}\rho
N\left( N-1\right) \left( 2N-1\right) \left( 1-\frac{5}{4}\delta \right) },
\label{BMWZBq1} \\
\mathrm{BMZB}_{q_{2}^{\alpha }}\left( \theta \right)  &=&\frac{\frac{\Gamma
\left( \alpha \right) ^{4}}{\Gamma \left( 2\alpha \right) ^{2}}}{2\left(
\alpha -1\right) ^{2}\frac{\Gamma \left( 2\alpha -3\right) \Gamma \left(
2\alpha -1\right) -\Gamma \left( 2\left( \alpha -1\right) \right) ^{2}}{%
\Gamma \left( 4\left( \alpha -1\right) \right) }+\frac{4\pi ^{2}\rho N\left(
N-1\right) \left( 2N-1\right) }{3}\frac{\Gamma \left( 2\alpha -1\right) ^{2}%
}{\Gamma \left( 2\left( 2\alpha -1\right) \right) }},  \label{BMWZBq2}
\end{eqnarray}%
where $\Gamma \left( \alpha \right) =\tint\nolimits_{0}^{\infty }x^{\alpha
-1}e^{-x}dx$ is the gamma function, and $\mathrm{WWB}_{q}$ (\ref{MOD WWB}-%
\ref{MOD WWB(h,s,q)}) becomes:
\end{subequations}
\begin{subequations}
\begin{equation}
\mathrm{WWB}_{q}\left( \theta \right) =\underset{s\in ]0,1[,|h|<1}{\sup }%
\left\{ \mathrm{WWB}_{q}^{h,s}\left( \theta \right) \right\} ,
\label{MOD WWB ST}
\end{equation}%
\begin{equation}
\mathrm{WWB}_{q}^{h,s}\left( \theta \right) =\left\{
\begin{array}{cc}
\frac{h^{2}\left( \tint\limits_{h}^{1}q(\theta )d\theta \right)
^{2}e^{4\left( s-1\right) s\rho v\left( h\right) }}{\left( e^{4s\left(
2s-1\right) \rho v\left( h\right) }+e^{4\left( s-1\right) \left( 2s-1\right)
\rho v\left( h\right) }\right) \tint\limits_{h}^{1}q(\theta )^{2}d\theta
-2e^{2s\left( s-1\right) \rho v\left( 2h\right)
}\tint\limits_{h}^{1-h}q(\theta )q\left( \theta +h\right) d\theta }, & \text{%
if }h\in \left[ 0,1\right[  \\
\frac{h^{2}\left( \tint\limits_{0}^{1+h}q(\theta )d\theta \right)
^{2}e^{4\left( s-1\right) s\rho v\left( h\right) }}{\left( e^{4s\left(
2s-1\right) \rho v\left( h\right) }+e^{4\left( s-1\right) \left( 2s-1\right)
\rho v\left( h\right) }\right) \tint\limits_{0}^{1+h}q(\theta )^{2}d\theta
-2e^{2s\left( s-1\right) \rho v\left( 2h\right)
}\tint\limits_{-h}^{1+h}q(\theta )q\left( \theta +h\right) d\theta }, &
\text{if }h\in \left] -1,0\right[
\end{array}%
\right. ,  \label{MOD WWB(h,s,q) ST}
\end{equation}%
where $\rho =\frac{\alpha ^{2}}{\sigma _{\mathbf{n}}^{2}}$ denotes the
(input) SNR and:
\begin{equation}
v\left( h\right) =%
\begin{cases}
N\left( 1-\cos \left( \pi \left( N-1\right) h\right) \frac{\sin \left( \pi
Nh\right) }{N\sin \left( \pi h\right) }\right) , & \text{if }h\neq 0 \\
0, & \text{if }h=0%
\end{cases}%
.
\end{equation}

\subsection{Comparisons and analysis for the single tone estimation}

First, we can derive from (\ref{BMWZBq ST}) an upper bound for $\mathrm{BCRB}%
\left( \theta \right) $ (\ref{BCRB ST}) in the asymptotic region. Indeed, in
the case of single tone estimation:
\end{subequations}
\begin{equation}
\mathrm{BMZB}_{q_{l}}\left( g\left( \theta \right) \right) =\frac{E_{\theta }%
\left[ \frac{dg\left( \theta \right) }{d\theta }q_{l}\left( \theta \right) %
\right] ^{2}}{E_{\theta }\left[ \left( \frac{\partial q_{l}\left( \theta
\right) }{\partial \theta }\right) ^{2}\right] +E_{\theta }\left[ c\left(
\theta \right) q_{l}\left( \theta \right) ^{2}\right] }\leq \frac{1}{c}\frac{%
E_{\theta }\left[ q_{l}\left( \theta \right) \right] ^{2}}{E_{\theta }\left[
q_{l}\left( \theta \right) ^{2}\right] }\leq \frac{1}{c},
\end{equation}%
where $c=\frac{4\pi ^{2}(N-1)(2N-1)}{3}N\rho $. Thus:
\begin{equation}
\mathrm{BCRB}\left( \theta \right) \underset{N\rho \rightarrow \infty }{%
\approx }\mathrm{BMZB}_{UB}\left( \theta \right) =\frac{3}{4\pi
^{2}(N-1)(2N-1)}\frac{1}{N\rho }.  \label{BCRB asymptotic ST}
\end{equation}%
Moreover, an upper bound on the minimum MSE, and therefore on any lower
bound on the MSE, is:%
\begin{equation}
\sigma _{\theta }^{2}=E_{\theta }\left[ \theta ^{2}\right] -E_{\theta }\left[
\theta \right] ^{2}=\frac{1}{12},  \label{upper bound ST}
\end{equation}%
which is also the MSE of the maximum a posteriori (MAP) estimator (which
coincides with the maximum likelihood estimate for uniform prior) in the
no-information region \cite{Van Trees - Bell}. \newline
As shown in figure (\ref{F: Fig1})\footnote{%
In all figures, $\mathrm{WWB}\left( \theta \right) $ (\ref{WWB ST 1-0}) and $%
\mathrm{WWB}_{q}\left( \theta \right) $ (\ref{MOD WWB ST}) are the supremum
computed over $h=-1+k10^{-4},1\leq k\leq 1999$, and $s=l10^{-2},1\leq l\leq
99$.}, the WWB (\ref{WWB ST 1-0}-\ref{WWB ST 1-1}) and the BZB (WWB where $%
s\rightarrow 1^{-}$) coincide with the $\mathrm{BMZB}_{UB}\left( \theta
\right) $ (\ref{BCRB asymptotic ST}) in the asymptotic region (where the WWB
and the BZB coincide with the MSE of the MAP \cite[pp 41-43]{Van Trees -
Bell}), although its limiting form $\mathrm{BCRB}_{1_{\mathcal{S}_{\mathcal{X%
},\Theta }}}\left( \theta \right) $ (\ref{WWB limiting form}) is zero.
\begin{figure}[tbph]
\centering\includegraphics[width=1\columnwidth,draft=false]{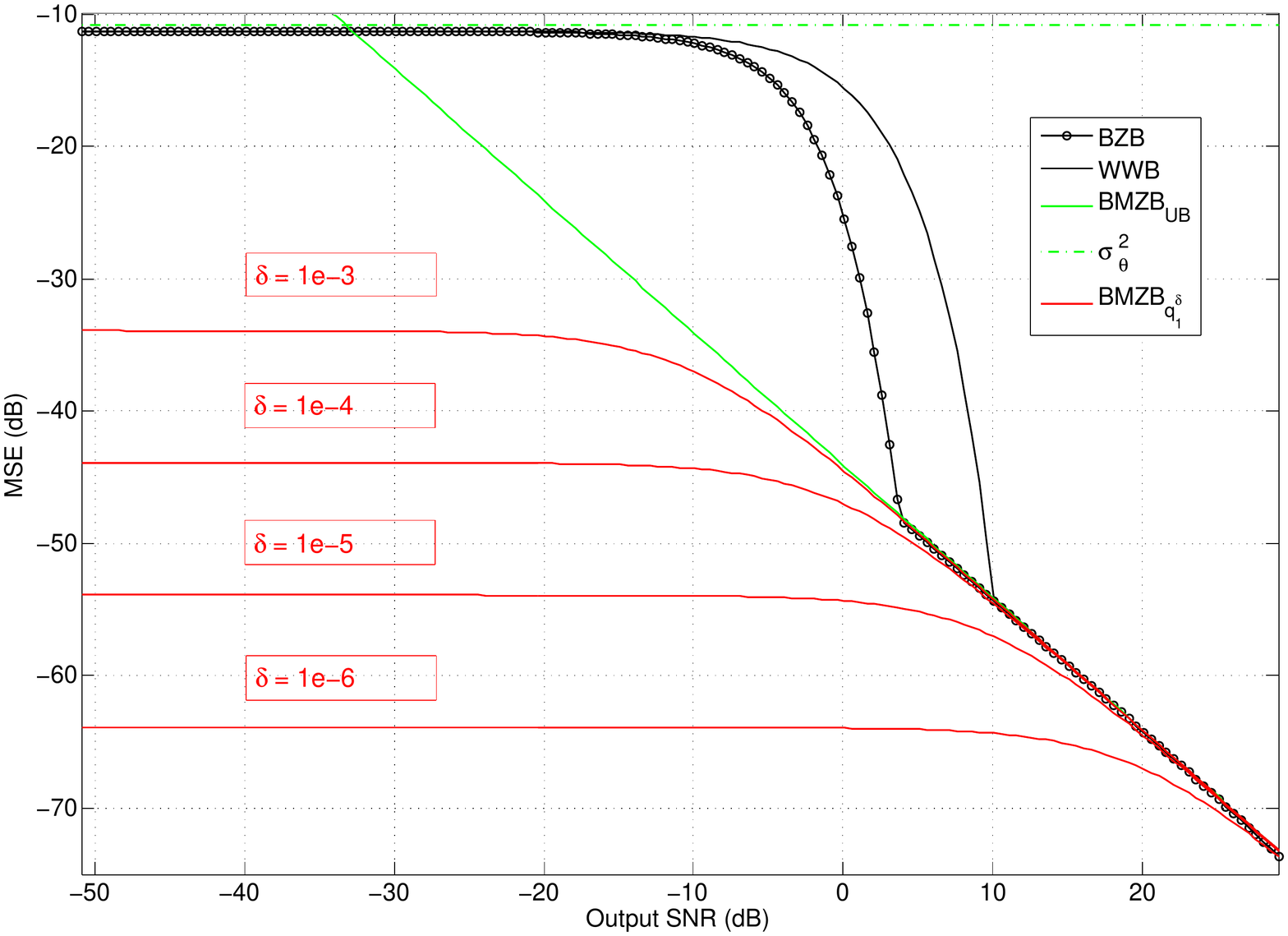}
\caption{Single tone estimation: illustration of the relationship between $%
\mathrm{BZB}\left( \protect\theta \right) $, $\mathrm{WWB}\left( \protect%
\theta \right) $ (\protect\ref{WWB ST 1-0}) and $\mathrm{BMZB}_{q_{1}^{%
\protect\delta }}\left( \protect\theta \right) $ (\protect\ref{BMWZBq1}), $%
N=32$, SNR step is 0.5 dB. }
\label{F: Fig1}
\end{figure}
Actually, this paradox can be explained by the fact that (\ref{limit
WWB(h,s)}):
\begin{subequations}
\begin{equation}
\mathrm{BCRB}_{1_{\mathcal{S}_{\mathcal{X},\Theta }}}\left( \theta \right) =%
\underset{h\rightarrow 0}{\lim }\mathrm{BZB}^{h}\left( \theta \right) =%
\underset{h\rightarrow 0}{\lim }\left( \underset{s\rightarrow 1^{-}}{\lim }%
\mathrm{WWB}^{h,s}\left( \theta \right) \right) =\underset{h\rightarrow 0}{%
\lim }\frac{1}{\frac{2}{\left\vert h\right\vert }+\frac{4}{3}\pi ^{2}\rho
N\left( N-1\right) \left( 2N-1\right) },
\end{equation}%
is similar to:%
\begin{equation}
\underset{\delta \rightarrow 0}{\lim }\mathrm{BMZB}_{q_{1}^{\delta }}\left(
\theta \right) =\frac{\left( 1-\delta \right) ^{2}}{\frac{\pi ^{2}}{4\delta }%
+\frac{4}{3}\pi ^{2}\rho N\left( N-1\right) \left( 2N-1\right) \left( 1-%
\frac{5}{4}\delta \right) }=\underset{\delta \rightarrow 0}{\lim }\frac{1}{%
\frac{\pi ^{2}}{4\delta }+\frac{4}{3}\pi ^{2}\rho N\left( N-1\right) \left(
2N-1\right) }
\end{equation}%
provided that, for any $\delta \ll 1$ one chooses $h\ll 1$ satisfying $%
\left\vert h\right\vert =8\delta /\pi ^{2}$.
\begin{figure}[tbp]
\centering\includegraphics[width=1\columnwidth,draft=false]{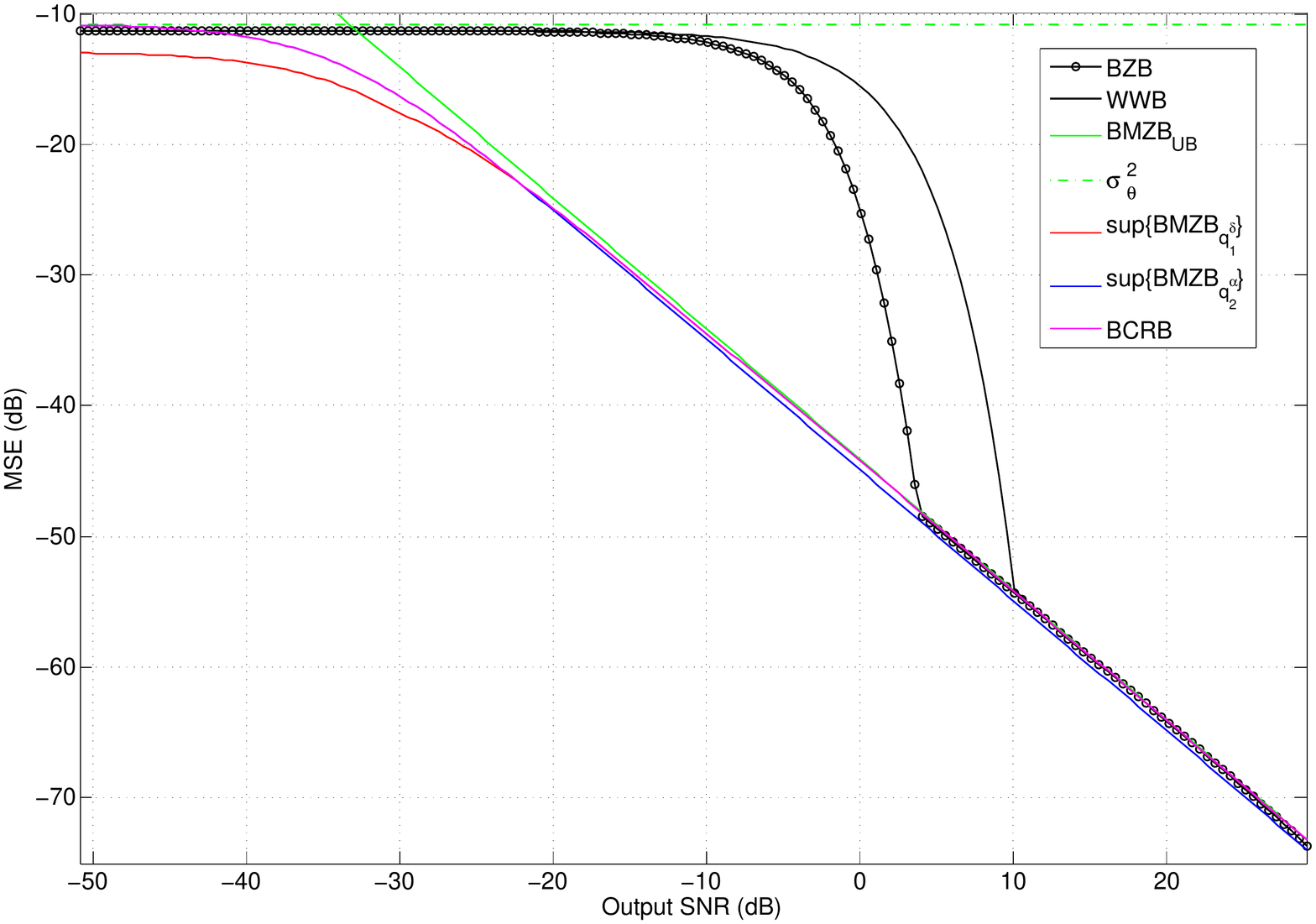}
\caption{Single tone estimation: comparison of some BCRB (\protect\ref{BCRB
ST}) obtained from two families of $\mathrm{BMZB}_{q}\left( \protect\theta %
\right) $ (\protect\ref{BMWZBq1}-\protect\ref{BMWZBq2}), $N=32$, SNR step is
0.5 dB.}
\label{F: Fig2}
\end{figure}
Therefore the limiting behavior of $\mathrm{BZB}^{h}\left( \theta \right) $
and $\mathrm{WWB}^{h,s}\left( \theta \right) $, where $h\rightarrow 0$, is
the limiting behavior of $\mathrm{BMZB}_{q_{1}^{\delta }}\left( \theta
\right) $, where $\delta \rightarrow 0$, which is exemplified in figure (\ref%
{F: Fig1}) as well, for $\delta \in \left\{
10^{-3},10^{-4},10^{-5},10^{-6}\right\} $. As mentioned above, $\mathrm{BMZB}%
_{q_{1}^{\delta }}\left( \theta \right) $ always yields asymptotically $%
\mathrm{BMZB}_{UB}\left( \theta \right) $ but is also upper bounded by:
\end{subequations}
\begin{equation}
\mathrm{BMZB}_{q_{1}^{\delta }}\left( \theta \right) \leq \frac{4\delta }{%
\pi ^{2}}
\end{equation}%
which tends to 0 when $\delta \rightarrow 0$. However, this adverse
numerical behaviour can be easily circumvented by resorting to $\mathrm{BMZB}%
_{q_{1}^{\delta }}\left( \theta \right) $ and a tight BCRB in the asymptotic
region can be obtained as $\underset{0<\delta \leq 0.5}{\sup }\left\{
\mathrm{BMZB}_{q_{1}^{\delta }}\left( \theta \right) \right\} $ (\ref{BCRB
ST}), as shown in figure (\ref{F: Fig2}). It is also worth noting that some
families of $\mathrm{BMZB}_{q}\left( \theta \right) $ does not allow to
obtain a tight BCRB in the asymptotic region, as already mentioned in \cite[%
pp 36-37]{Van Trees - Bell}, and again exemplified in the studied case in
figure (\ref{F: Fig2}), if we consider $\underset{\alpha \geq \frac{3}{2}}{%
\sup }\left\{ \mathrm{BMZB}_{q_{2}^{\alpha }}\left( \theta \right) \right\} $%
, which is however tight in the no-information region. Of course, one can
combine the two families of $\mathrm{BMZB}$ as in (\ref{BCRB ST}), in order
to obtain a BCRB tight both in the asymptotic and the no-information region:%
\begin{equation}
\mathrm{BCRB}\left( \theta \right) =\sup \left\{ \underset{0<\delta \leq 0.5}%
{\sup }\left\{ \mathrm{BMZB}_{q_{1}^{\delta }}\left( \theta \right) \right\}
,\underset{\alpha \geq \frac{3}{2}}{\sup }\left\{ \mathrm{BMZB}%
_{q_{2}^{\alpha }}\left( \theta \right) \right\} \right\} ,
\end{equation}%
as also shown in figure (\ref{F: Fig2}). \newline
Last, in figure (\ref{F: Fig3}) we display two different modified $\mathrm{%
WWB}_{q}\left( \theta \right) $ (\ref{MOD WWB ST}), namely the $\mathrm{WWB}%
_{q_{1}^{0.5}}\left( \theta \right) $ and the $\mathrm{WWB}%
_{q_{2}^{2}}\left( \theta \right) $, and the associated modified $\mathrm{BZB%
}_{q_{1}^{0.5}}\left( \theta \right) $ and $\mathrm{BZB}_{q_{2}^{2}}\left(
\theta \right) ,$ for a comparison with the $\mathrm{WWB}\left( \theta
\right) $ (\ref{WWB ST 1-0}) and the associated $\mathrm{BZB}\left( \theta
\right) $. Figure (\ref{F: Fig3}) highlights the following result: if the
\textit{non zero} limiting form of a large-error bound is tighter in the
asymptotic region than the \textit{non zero} limiting form of another
large-error bound, this tightness relationship is still valid in the
threshold region for the two large-error bounds.
\begin{figure}[tbp]
\centering\includegraphics[width=1\columnwidth,draft=false]{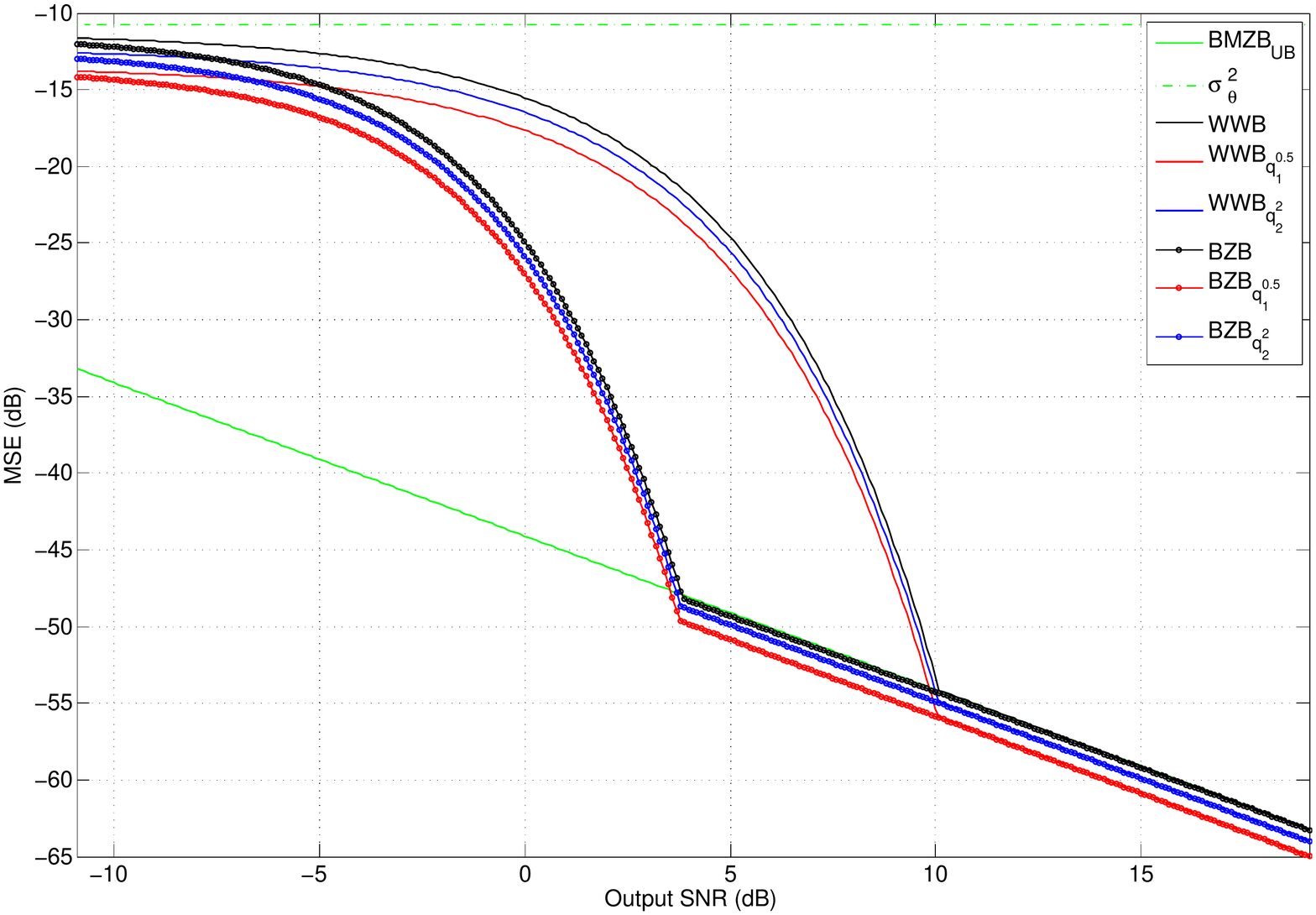}
\caption{Single tone estimation: comparision of various modified BZBs and
WWBs (\protect\ref{MOD WWB ST}), $N=32$, SNR step is 0.1 dB.}
\label{F: Fig3}
\end{figure}
Although not displayed, we have checked this result in all the numerous
comparisons we have done between representatives of $\mathrm{BMZB}%
_{q_{1}^{\delta }}\left( \theta \right) $ and $\mathrm{BMZB}_{q_{2}^{\alpha
}}\left( \theta \right) $, within the same family or not. More precisely, we
have noticed that for both the WWB and the BZB, the threshold value does not
change (with a precision of 0.1 dB), but the relative bound tightness in the
threshold region depends on the relative bound tightness in the asymptotic
region. This observation allows to understand why the $\mathrm{WWB}\left(
\theta \right) $ and the $\mathrm{BZB}\left( \theta \right) $ remain the
tightest bounds in the threshold region. Indeed as:%
\begin{equation}
\mathrm{WWB}\left( \theta \right) =\underset{\delta \rightarrow 0}{\lim }%
\left\{ \mathrm{WWB}_{q_{1}^{\delta }}\left( \theta \right) \right\} \text{
and }\mathrm{BZB}\left( \theta \right) =\underset{\delta \rightarrow 0}{\lim
}\left\{ \mathrm{BZB}_{q_{1}^{\delta }}\left( \theta \right) \right\} ,
\end{equation}%
therefore, asymptotically, the limiting form of both the $\mathrm{WWB}\left(
\theta \right) $ and the $\mathrm{BZB}\left( \theta \right) $ is $\mathrm{%
BMZB}_{q_{1}^{\delta }}\left( \theta \right) $, which asymptotically
coincides with $\mathrm{BMZB}_{UB}\left( \theta \right) $, the tightest
value of $\mathrm{BMZB}_{q}\left( \theta \right) $.

\section{Conclusion}

In the present paper, a fairly general class of "large-error" BLBs of the
WWF, essentially free from regularity conditions on the pdfs support and for
which a limiting form yields a generalized BCRB, has been introduced. The
proposed class of BLBs defines a wide range of Bayesian estimation problems
for which a non trivial generalized BCRB exists, which is a key result from
a practical viewpoint. In a large number of cases, this limiting form
appears to be the BMZB. This theoretical result open new perspectives in the
search of tight lower-bounds in the threshold region, new ones or some
modified existing ones. Indeed, since the BMZB may provide a tighter bound
than the historical BCRB in the asymptotic region, modified WWB and BZB
which limiting form is the BMZB has been proposed. The analysis of the
behavior of the proposed modified bounds in an application case has led us
to postulate the following conjecture: if the \textit{non zero} limiting
form of a large-error bound is tighter in the asymptotic region than the
\textit{non zero} limiting form of another large-error bound, this tightness
relationship is still valid in the threshold region for the two large-error
bounds. Further study cases need to be addressed in order to quantify how
general or specific is this conjecture.

\section{Appendix}

In this Appendix, Propositions 1 and its extension, Proposition 2, are
derived.

\subsection{Case of bounded intervals\label{A: Case of bounded intervals}}

First, let us address the case of closed intervals:
\begin{equation}
\forall \mathbf{x}\in \mathcal{S}_{\mathcal{X}},\quad \mathcal{S}_{\Theta |%
\mathbf{x}}=\left[ a_{\mathbf{x}},b_{\mathbf{x}}\right] :-\infty <a_{\mathbf{%
x}}<b_{\mathbf{x}}<+\infty .  \label{A - connected subset case}
\end{equation}%
\textbf{\smallskip Step 1}\smallskip \newline
First, one needs to asses:
\begin{equation}
\lim_{h\rightarrow 0}E_{\mathbf{x},\theta }\left[ g\left( \theta \right)
\frac{\psi _{q}^{h}\left( \mathbf{x},\theta \right) }{h}\right]
=\lim_{h\rightarrow 0}E_{\mathbf{x},\theta }\left[ \frac{g\left( \theta
-h\right) -g\left( \theta \right) }{h}q\left( \mathbf{x},\theta \right) 1_{%
\mathcal{S}_{\Theta |\mathbf{x}}}\left( \theta -h\right) \right]
\end{equation}%
According to (\ref{A - connected subset case}), $\forall \mathbf{x}\in
\mathcal{S}_{\mathcal{X}},\forall h>0$:
\begin{subequations}
\begin{equation}
\begin{array}{c}
E_{\theta |\mathbf{x}}\left[ \frac{g\left( \theta -h\right) -g\left( \theta
\right) }{h}q\left( \mathbf{x},\theta \right) 1_{\mathcal{S}_{\Theta |%
\mathbf{x}}}\left( \theta -h\right) \right] =-\dint\limits_{a_{\mathbf{x}%
}+h}^{b_{\mathbf{x}}}\left( \frac{g\left( \theta -h\right) -g\left( \theta
\right) }{-h}\right) q\left( \mathbf{x},\theta \right) p\left( \theta |%
\mathbf{x}\right) d\theta \\
E_{\theta |\mathbf{x}}\left[ \frac{g\left( \theta +h\right) -g\left( \theta
\right) }{-h}q\left( \mathbf{x},\theta \right) 1_{\mathcal{S}_{\Theta |%
\mathbf{x}}}\left( \theta +h\right) \right] =-\dint\limits_{a_{\mathbf{x}%
}}^{b_{\mathbf{x}}-h}\left( \frac{g\left( \theta +h\right) -g\left( \theta
\right) }{h}\right) q\left( \mathbf{x},\theta \right) p\left( \theta |%
\mathbf{x}\right) d\theta%
\end{array}%
\quad
\end{equation}%
Assuming that $g\left( \theta \right) $ is of class $\mathcal{C}^{1}$ over $%
\mathcal{S}_{\Theta |\mathbf{x}}$, by invoking the mean value theorem \cite%
{Orsay}, one obtains
\begin{equation}
\begin{array}{c}
\exists \gamma ^{+}\left( h\right) \in \left[ 0,h\right] :\frac{g\left(
\theta +h\right) -g\left( \theta \right) }{h}=\frac{dg\left( \theta +\gamma
^{+}\left( h\right) \right) }{d\theta } \\
\exists \gamma ^{-}\left( h\right) \in \left[ 0,h\right] :\frac{g\left(
\theta \right) -g\left( \theta -h\right) }{h}=\frac{dg\left( \theta -\gamma
^{-}\left( h\right) \right) }{d\theta }%
\end{array}%
\end{equation}%
Thus, we deduce that, $\forall \mathbf{x}\in \mathcal{S}_{\mathcal{X}}:$
\begin{eqnarray}
&&%
\begin{array}{l}
\lim\limits_{h\rightarrow 0^{+}}E_{\theta |\mathbf{x}}\left[ \frac{g\left(
\theta -h\right) -g\left( \theta \right) }{h}q\left( \mathbf{x},\theta
\right) 1_{\mathcal{S}_{\Theta |\mathbf{x}}}\left( \theta -h\right) \right]
=-\lim\limits_{h\rightarrow 0^{+}}\dint\limits_{a_{\mathbf{x}}+h}^{b_{%
\mathbf{x}}}\frac{dg\left( \theta \right) }{d\theta }q\left( \mathbf{x}%
,\theta \right) p\left( \theta |\mathbf{x}\right) d\theta \qquad \qquad
\qquad \qquad \\
\qquad \qquad \qquad \qquad \qquad \qquad \qquad \qquad
~+\lim\limits_{h\rightarrow 0^{+}}\dint\limits_{a_{\mathbf{x}}+h}^{b_{%
\mathbf{x}}}\left( \frac{dg\left( \theta \right) }{d\theta }-\frac{dg\left(
\theta -\gamma ^{-}\left( h\right) \right) }{d\theta }\right) q\left(
\mathbf{x},\theta \right) p\left( \theta |\mathbf{x}\right) d\theta%
\end{array}
\\
&&%
\begin{array}{l}
\lim\limits_{h\rightarrow 0^{+}}E_{\theta |\mathbf{x}}\left[ \frac{g\left(
\theta +h\right) -g\left( \theta \right) }{-h}q\left( \mathbf{x},\theta
\right) 1_{\mathcal{S}_{\Theta |\mathbf{x}}}\left( \theta +h\right) \right]
=-\lim\limits_{h\rightarrow 0^{+}}\dint\limits_{a_{\mathbf{x}}}^{b_{\mathbf{x%
}}-h}\frac{dg\left( \theta \right) }{d\theta }q\left( \mathbf{x},\theta
\right) p\left( \theta |\mathbf{x}\right) d\theta \qquad \qquad \qquad \qquad
\\
\qquad \qquad \qquad \qquad \qquad \qquad \qquad \qquad
~+\lim\limits_{h\rightarrow 0^{+}}\dint\limits_{a_{\mathbf{x}}}^{b_{\mathbf{x%
}}-h}\left( \frac{dg\left( \theta \right) }{d\theta }-\frac{dg\left( \theta
+\gamma ^{+}\left( h\right) \right) }{d\theta }\right) q\left( \mathbf{x}%
,\theta \right) p\left( \theta |\mathbf{x}\right) d\theta%
\end{array}%
\end{eqnarray}%
According to Heine theorem \cite{Orsay}:
\end{subequations}
\begin{subequations}
\begin{equation}
\forall \varepsilon >0,\exists h>0~|~\forall \left( \theta ,\theta ^{\prime
}\right) \in \mathcal{S}_{\theta |\mathbf{x}},\left\vert \theta -\theta
^{\prime }\right\vert <h~\Rightarrow ~\left\vert \frac{dg\left( \theta
\right) }{d\theta }-\frac{dg\left( \theta ^{\prime }\right) }{d\theta }%
\right\vert <\varepsilon ,  \label{A - Heine}
\end{equation}%
one can state that
\begin{equation}
\forall \varepsilon >0,\exists h>0~|~\left\{
\begin{array}{c}
\forall \gamma ^{+}\left( h\right) <h~\Rightarrow ~\left\vert \frac{dg\left(
\theta \right) }{d\theta }-\frac{dg\left( \theta +\gamma ^{+}\left( h\right)
\right) }{d\theta }\right\vert <\varepsilon \\
\forall \gamma ^{-}\left( h\right) <h~\Rightarrow ~\left\vert \frac{dg\left(
\theta \right) }{d\theta }-\frac{dg\left( \theta -\gamma ^{-}\left( h\right)
\right) }{d\theta }\right\vert <\varepsilon%
\end{array}%
\right. .
\end{equation}%
Consequently,$\forall \mathbf{x}\in \mathcal{S}_{\mathcal{X}}$, $\forall
\varepsilon >0,\exists h>0$ such that:
\begin{eqnarray}
\left\vert \dint\limits_{a_{\mathbf{x}}+h}^{b_{\mathbf{x}}}\left(
\begin{array}{l}
\frac{dg\left( \theta \right) }{d\theta }- \\
\frac{dg\left( \theta -\gamma ^{-}\left( h\right) \right) }{d\theta }%
\end{array}%
\right) q\left( \mathbf{x},\theta \right) p\left( \theta |\mathbf{x}\right)
d\theta \right\vert &\leq &\dint\limits_{a_{\mathbf{x}}+h}^{b_{\mathbf{x}%
}}\left\vert \frac{dg\left( \theta \right) }{d\theta }-\frac{dg\left( \theta
-\gamma ^{-}\left( h\right) \right) }{d\theta }\right\vert \left\vert
q\left( \mathbf{x},\theta \right) \right\vert p\left( \theta |\mathbf{x}%
\right) d\theta \qquad \quad \\
&\leq &\dint\limits_{a_{\mathbf{x}}}^{b_{\mathbf{x}}}\left\vert \frac{%
dg\left( \theta \right) }{d\theta }-\frac{dg\left( \theta -\gamma ^{-}\left(
h\right) \right) }{d\theta }\right\vert \left\vert q\left( \mathbf{x},\theta
\right) \right\vert p\left( \theta |\mathbf{x}\right) d\theta \\
&<&\varepsilon E_{\theta |\mathbf{x}}\left[ \left\vert q\left( \mathbf{x}%
,\theta \right) \right\vert \right]
\end{eqnarray}%
leading to:
\end{subequations}
\begin{subequations}
\begin{equation}
\forall \mathbf{x}\in \mathcal{S}_{\mathcal{X}}:\lim_{h\rightarrow
0}E_{\theta |\mathbf{x}}\left[ g\left( \theta \right) \frac{\psi
_{q}^{h}\left( \mathbf{x},\theta \right) }{h}\right] =-E_{\theta |\mathbf{x}}%
\left[ \frac{\partial g\left( \theta \right) }{\partial \theta }q\left(
\mathbf{x},\theta \right) \right] ,
\end{equation}%
and:%
\begin{equation}
\lim_{h\rightarrow 0}E_{\mathbf{x},\theta }\left[ g\left( \theta \right)
\frac{\psi _{q}^{h}\left( \mathbf{x},\theta \right) }{h}\right] =-E_{\mathbf{%
x},\theta }\left[ \frac{\partial g\left( \theta \right) }{\partial \theta }%
q\left( \mathbf{x},\theta \right) \right] .  \label{A - ident 0}
\end{equation}%
\newline
\textbf{Step 2}\smallskip \newline
Second, one needs to asses:
\end{subequations}
\begin{equation}
\lim_{h\rightarrow 0}E_{\mathbf{x},\theta }\left[ \left( \frac{\psi
_{q}^{h}\left( \mathbf{x},\theta \right) }{h}\right) ^{2}\right] .
\end{equation}%
According to (\ref{A - connected subset case}), $\forall \mathbf{x}\in
\mathcal{S}_{\mathcal{X}},\forall h>0$:
\begin{subequations}
\begin{eqnarray}
E_{\theta |\mathbf{x}}\left[ \frac{\psi _{q}^{h}\left( \mathbf{x},\theta
\right) ^{2}}{h^{2}}\right] &=&\frac{1}{h^{2}}\dint\limits_{a_{\mathbf{x}%
}}^{b_{\mathbf{x}}}\left(
\begin{array}{l}
q\left( \mathbf{x},\theta +h\right) p\left( \theta +h|\mathbf{x}\right) 1_{%
\mathcal{S}_{\theta {|\mathbf{x}}}}\left( \theta +h\right) \\
-q\left( \mathbf{x},\theta \right) p\left( \theta |\mathbf{x}\right) 1_{%
\mathcal{S}_{\Theta |\mathbf{x}}}\left( \theta -h\right)%
\end{array}%
\right) ^{2}\frac{d\theta }{p\left( \theta |\mathbf{x}\right) } \\
E_{\theta |\mathbf{x}}\left[ \frac{\psi _{q}^{h}\left( \mathbf{x},\theta
\right) ^{2}}{h^{2}}\right] &=&%
\begin{array}{l}
\dint\limits_{a_{\mathbf{x}}}^{b_{\mathbf{x}}-h}\frac{\left( q\left( \mathbf{%
x},\theta +h\right) p\left( \theta +h|\mathbf{x}\right) -q\left( \mathbf{x}%
,\theta \right) p\left( \theta |\mathbf{x}\right) \right) ^{2}}{h^{2}}\frac{%
d\theta }{p\left( \theta |\mathbf{x}\right) } \\
+\frac{2}{h}\dint\limits_{a_{\mathbf{x}}}^{a_{\mathbf{x}}+h}\frac{q\left(
\mathbf{x},\theta +h\right) p\left( \theta +h|\mathbf{x}\right) -q\left(
\mathbf{x},\theta \right) p\left( \theta |\mathbf{x}\right) }{h}q\left(
\mathbf{x},\theta \right) d\theta \\
+\frac{1}{h^{2}}\left( \dint\limits_{a_{\mathbf{x}}}^{a_{\mathbf{x}%
}+h}q\left( \mathbf{x},\theta \right) ^{2}p\left( \theta |\mathbf{x}\right)
d\theta +\dint\limits_{b_{\mathbf{x}}-h}^{b_{\mathbf{x}}}q\left( \mathbf{x}%
,\theta \right) ^{2}p\left( \theta |\mathbf{x}\right) d\theta \right)%
\end{array}%
\end{eqnarray}%
and%
\begin{eqnarray}
E_{\theta |\mathbf{x}}\left[ \frac{\psi _{q}^{-h}\left( \mathbf{x},\theta
\right) ^{2}}{\left( -h\right) ^{2}}\right] &=&\frac{1}{h^{2}}%
\dint\limits_{a_{\mathbf{x}}}^{b_{\mathbf{x}}}\left(
\begin{array}{l}
q\left( \mathbf{x},\theta -h\right) p\left( \theta -h|\mathbf{x}\right) 1_{%
\mathcal{S}_{\theta |\mathbf{x}}}\left( \theta -h\right) \\
-q\left( \mathbf{x},\theta \right) p\left( \theta |\mathbf{x}\right) 1_{%
\mathcal{S}_{\theta |\mathbf{x}}}\left( \theta +h\right)%
\end{array}%
\right) ^{2}\frac{d\theta }{p\left( \theta |\mathbf{x}\right) } \\
E_{\theta |\mathbf{x}}\left[ \frac{\psi _{q}^{-h}\left( \mathbf{x},\theta
\right) ^{2}}{\left( -h\right) ^{2}}\right] &=&%
\begin{array}{l}
\dint\limits_{a_{\mathbf{x}}+h}^{b_{\mathbf{x}}}\frac{\left( q\left( \mathbf{%
x},\theta +h\right) p\left( \theta +h|\mathbf{x}\right) -q\left( \mathbf{x}%
,\theta \right) p\left( \theta |\mathbf{x}\right) \right) ^{2}}{\left(
-h\right) ^{2}}\frac{d\theta }{p\left( \theta |\mathbf{x}\right) } \\
-\frac{2}{h}\dint\limits_{b_{\mathbf{x}}-h}^{b_{\mathbf{x}}}\frac{q\left(
\mathbf{x},\theta -h\right) p\left( \theta -h|\mathbf{x}\right) -q\left(
\mathbf{x},\theta \right) p\left( \theta |\mathbf{x}\right) }{-h}q\left(
\mathbf{x},\theta \right) d\theta \\
+\frac{1}{h^{2}}\left( \dint\limits_{a_{\mathbf{x}}}^{a_{\mathbf{x}%
}+h}q\left( \mathbf{x},\theta \right) ^{2}p\left( \theta |\mathbf{x}\right)
d\theta +\dint\limits_{b_{\mathbf{x}}-h}^{b_{\mathbf{x}}}q\left( \mathbf{x}%
,\theta \right) ^{2}p\left( \theta |\mathbf{x}\right) d\theta \right)%
\end{array}%
\end{eqnarray}

Assuming that $t\left( \mathbf{x},\theta \right) =q\left( \mathbf{x},\theta
\right) p\left( \theta |\mathbf{x}\right) $ is of class $\mathcal{C}^{1}$
w.r.t. $\theta $ over $\mathcal{S}_{\Theta |\mathbf{x}}$, thus $\left( \frac{%
\partial t\left( \mathbf{x},\theta \right) }{\partial \theta }\right) ^{2}$
is continuous over $\mathcal{S}_{\Theta |\mathbf{x}}$. Then, using the same
rationale as in step 1 based on the mean value theorem and the Heine
theorem, one can easily prove that, $\forall \mathbf{x}\in \mathcal{S}_{%
\mathcal{X}}$:

\begin{equation}
\lim\limits_{h\rightarrow 0^{+}}E_{\theta |\mathbf{x}}\left[ \left( \frac{%
\psi _{q}^{h}\left( \mathbf{x},\theta \right) }{h}\right) ^{2}\right] =%
\begin{array}{l}
E_{\theta |\mathbf{x}}\left[ \left( \frac{1}{p\left( \theta |\mathbf{x}%
\right) }\frac{\partial t\left( \mathbf{x},\theta \right) }{\partial \theta }%
\right) ^{2}\right] +2q\left( \mathbf{x},a_{\mathbf{x}}\right) \left. \frac{%
\partial t\left( \mathbf{x},\theta \right) }{\partial \theta }\right\vert
_{\theta =a_{\mathbf{x}}} \\
+\frac{1}{h^{2}}\left( \dint\limits_{a_{\mathbf{x}}}^{a_{\mathbf{x}%
}+h}u\left( \mathbf{x},\theta \right) d\theta +\dint\limits_{b_{\mathbf{x}%
}-h}^{b_{\mathbf{x}}}u\left( \mathbf{x},\theta \right) d\theta \right)
\end{array}
\label{A - ident 1 - h+}
\end{equation}%
and
\begin{equation}
\lim\limits_{h\rightarrow 0^{+}}E_{\theta |\mathbf{x}}\left[ \left( \frac{%
\psi _{q}^{-h}\left( \mathbf{x},\theta \right) }{-h}\right) ^{2}\right] =%
\begin{array}{l}
E_{\theta |\mathbf{x}}\left[ \left( \frac{1}{p\left( \theta |\mathbf{x}%
\right) }\frac{\partial t\left( \mathbf{x},\theta \right) }{\partial \theta }%
\right) ^{2}\right] -2q\left( \mathbf{x},b_{\mathbf{x}}\right) \left. \frac{%
\partial t\left( \mathbf{x},\theta \right) }{\partial \theta }\right\vert
_{\theta =b_{\mathbf{x}}} \\
+\frac{1}{h^{2}}\left( \dint\limits_{a_{\mathbf{x}}}^{a_{\mathbf{x}%
}+h}u\left( \mathbf{x},\theta \right) d\theta +\dint\limits_{b_{\mathbf{x}%
}-h}^{b_{\mathbf{x}}}u\left( \mathbf{x},\theta \right) d\theta \right)
\end{array}
\label{A - ident 1 - h-}
\end{equation}%
where $u\left( \mathbf{x},\theta \right) =t\left( \mathbf{x},\theta \right)
q\left( \mathbf{x},\theta \right) =q\left( \mathbf{x},\theta \right)
^{2}p\left( \theta |\mathbf{x}\right) $. Assuming that $u\left( \mathbf{x}%
,\theta \right) $ is of class $\mathcal{C}^{2}$ w.r.t. $\theta $ at the
vicinity of endpoints $a_{\mathbf{x}}$ and $b_{\mathbf{x}}$, one can prove
that (see appendix \ref{A: limite 1}):
\end{subequations}
\begin{multline}
\lim_{h\rightarrow 0^{+}}\frac{1}{h^{2}}\left( \dint\limits_{a_{\mathbf{x}%
}}^{a_{\mathbf{x}}+h}u\left( \mathbf{x},\theta \right) d\theta
+\dint\limits_{b_{\mathbf{x}}-h}^{b_{\mathbf{x}}}u\left( \mathbf{x},\theta
\right) d\theta \right) =\frac{u\left( \mathbf{x},b_{\mathbf{x}}\right)
+u\left( \mathbf{x},a_{\mathbf{x}}\right) }{h}  \label{eq_annexe} \\
+\frac{1}{2}\left( \left. \frac{\partial u\left( \mathbf{x},{\theta }\right)
}{\partial \theta }\right\vert _{\theta =a_{\mathbf{x}}}-\left. \frac{%
\partial u\left( \mathbf{x},{\theta }\right) }{\partial \theta }\right\vert
_{\theta =b_{\mathbf{x}}}\right)
\end{multline}%
Therefore, in order to obtain a non trivial $\mathrm{BCRB}_{q}$ from (\ref{A
- ident 1 - h+}-\ref{A - ident 1 - h-}), the following necessary and
sufficient conditions must hold:
\begin{subequations}
\begin{equation}
u\left( \mathbf{x},b_{\mathbf{x}}\right) +u\left( \mathbf{x},a_{\mathbf{x}%
}\right) =q\left( \mathbf{x},a_{\mathbf{x}}\right) ^{2}p\left( a_{\mathbf{x}%
}|\mathbf{x}\right) +q\left( \mathbf{x},b_{\mathbf{x}}\right) ^{2}p\left( b_{%
\mathbf{x}}|\mathbf{x}\right) =0,
\end{equation}%
that is:
\begin{equation}
q\left( \mathbf{x},a_{\mathbf{x}}\right) p\left( a_{\mathbf{x}}|\mathbf{x}%
\right) =0\text{ and }q\left( \mathbf{x},b_{\mathbf{x}}\right) p\left( b_{%
\mathbf{x}}|\mathbf{x}\right) =0.  \label{condition_sff_nec}
\end{equation}%
Plugging (\ref{condition_sff_nec}) into the following identities
\begin{eqnarray}
\frac{\partial u\left( \mathbf{x},\theta \right) }{\partial \theta } &=&2%
\frac{\partial q\left( \mathbf{x},\theta \right) }{\partial \theta }q\left(
\mathbf{x},\theta \right) p\left( \theta |\mathbf{x}\right) +q\left( \mathbf{%
x},\theta \right) ^{2}\frac{\partial p\left( \theta |\mathbf{x}\right) }{%
\partial \theta }, \\
\frac{\partial t\left( \mathbf{x},\theta \right) }{\partial \theta }q\left(
\mathbf{x},\theta \right)  &=&\frac{\partial q\left( \mathbf{x},\theta
\right) }{\partial \theta }q\left( \mathbf{x},\theta \right) p\left( \theta |%
\mathbf{x}\right) +q\left( \mathbf{x},\theta \right) ^{2}\frac{\partial
p\left( \theta |\mathbf{x}\right) }{\partial \theta ^{2}},
\end{eqnarray}%
one obtains
\begin{eqnarray}
\left. \frac{\partial t\left( \mathbf{x},\theta \right) }{\partial \theta }%
\right\vert _{\theta =a_{\mathbf{x}}}q\left( \mathbf{x},a_{\mathbf{x}%
}\right)  &=&q\left( \mathbf{x},a_{\mathbf{x}}\right) ^{2}\left. \frac{%
\partial p\left( {\theta }|\mathbf{x}\right) }{\partial \theta }\right\vert
_{\theta =a_{\mathbf{x}}}=\left. \frac{\partial u\left( \mathbf{x},{\theta }%
\right) }{\partial \theta }\right\vert _{\theta =a_{\mathbf{x}}}, \\
\left. \frac{\partial t\left( \mathbf{x},\theta \right) }{\partial \theta }%
\right\vert _{\theta =b_{\mathbf{x}}}q\left( \mathbf{x},a_{\mathbf{x}%
}\right)  &=&q\left( \mathbf{x},b_{\mathbf{x}}\right) ^{2}\left. \frac{%
\partial p\left( {\theta }|\mathbf{x}\right) }{\partial \theta }\right\vert
_{\theta =b_{\mathbf{x}}}=\left. \frac{\partial u\left( \mathbf{x},\theta
\right) }{\partial \theta }\right\vert _{\theta =b_{\mathbf{x}}},
\end{eqnarray}%
leading to, $\forall \mathbf{x}\in \mathcal{S}_{\mathcal{X}}:$%
\end{subequations}
\begin{subequations}
\begin{eqnarray}
\lim\limits_{h\rightarrow 0^{+}}E_{\theta |\mathbf{x}}\left[ \left( \frac{%
\psi _{q}^{h}\left( \mathbf{x},\theta \right) }{h}\right) ^{2}\right]  &=&%
\begin{array}{l}
E_{\theta |\mathbf{x}}\left[ \left( \frac{1}{p\left( \theta |\mathbf{x}%
\right) }\frac{\partial t\left( \mathbf{x},\theta \right) }{\partial \theta }%
\right) ^{2}\right]  \\
+\frac{5}{2}q\left( \mathbf{x},a_{\mathbf{x}}\right) ^{2}\left. \frac{%
\partial p\left( {\theta }|\mathbf{x}\right) }{\partial \theta }\right\vert
_{\theta =a_{\mathbf{x}}}-\frac{1}{2}q\left( \mathbf{x},b_{\mathbf{x}%
}\right) ^{2}\left. \frac{\partial p\left( \theta |\mathbf{x}\right) }{%
\partial \theta }\right\vert _{\theta =b_{\mathbf{x}}}%
\end{array}
\label{A - ident 2 - h+} \\
\lim\limits_{h\rightarrow 0^{+}}E_{\theta |\mathbf{x}}\left[ \left( \frac{%
\psi _{q}^{-h}\left( \mathbf{x},\theta \right) }{-h}\right) ^{2}\right]  &=&%
\begin{array}{l}
E_{\theta |\mathbf{x}}\left[ \left( \frac{1}{p\left( \theta |\mathbf{x}%
\right) }\frac{\partial t\left( \mathbf{x},\theta \right) }{\partial \theta }%
\right) ^{2}\right]  \\
-\frac{5}{2}q\left( \mathbf{x},b_{\mathbf{x}}\right) ^{2}\left. \frac{%
\partial p\left( {\theta }|\mathbf{x}\right) }{\partial \theta }\right\vert
_{\theta =b_{\mathbf{x}}}+\frac{1}{2}q\left( \mathbf{x},a_{\mathbf{x}%
}\right) ^{2}\left. \frac{\partial p\left( {\theta }|\mathbf{x}\right) }{%
\partial \theta }\right\vert _{\theta =a_{\mathbf{x}}}%
\end{array}
\label{A - ident 2 - h-}
\end{eqnarray}%
Furthermore, the endpoints condition $u\left( \mathbf{x},a_{\mathbf{x}%
}\right) =0$ and $u\left( \mathbf{x},b_{\mathbf{x}}\right) =0$ implies that
the function $u\left( \mathbf{x},\theta \right) $ is increasing at the
vicinity of $a_{\mathbf{x}}$ and decreasing at the vicinity of $b_{\mathbf{x}%
}$. Thus:
\end{subequations}
\begin{subequations}
\begin{equation}
\frac{5}{2}q\left( \mathbf{x},a_{\mathbf{x}}\right) ^{2}\left. \frac{%
\partial p\left( {\theta }|\mathbf{x}\right) }{\partial \theta }\right\vert
_{\theta =a_{\mathbf{x}}}-\frac{1}{2}q\left( \mathbf{x},b_{\mathbf{x}%
}\right) ^{2}\left. \frac{\partial p\left( {\theta }|\mathbf{x}\right) }{%
\partial \theta }\right\vert _{\theta =b_{\mathbf{x}}}\geq 0
\end{equation}%
and
\begin{equation}
-\frac{5}{2}q\left( \mathbf{x},b_{\mathbf{x}}\right) ^{2}\left. \frac{%
\partial p\left( {\theta }|\mathbf{x}\right) }{\partial \theta }\right\vert
_{\theta =b_{\mathbf{x}}}+\frac{1}{2}q\left( \mathbf{x},a_{\mathbf{x}%
}\right) ^{2}\left. \frac{\partial p\left( {\theta }|\mathbf{x}\right) }{%
\partial \theta }\right\vert _{\theta =a_{\mathbf{x}}}\geq 0
\end{equation}%
Consequently:
\end{subequations}
\begin{subequations}
\begin{equation}
\lim\limits_{h\rightarrow 0}E_{\theta |\mathbf{x}}\left[ \left( \frac{\psi
_{q}^{h}\left( \mathbf{x},\theta \right) }{h}\right) ^{2}\right] \geq
E_{\theta |\mathbf{x}}\left[ \left( \frac{1}{p\left( \theta |\mathbf{x}%
\right) }\frac{\partial t\left( \mathbf{x},\theta \right) }{\partial \theta }%
\right) ^{2}\right]   \label{inequality BCRB 1}
\end{equation}%
in which, the equality holds for $q\left( \mathbf{x},a_{\mathbf{x}}\right)
=q\left( \mathbf{x},b_{\mathbf{x}}\right) =0$. Last, let $v\left( \mathbf{x}%
,\theta \right) =q\left( \mathbf{x},\theta \right) ^{2}\frac{\partial
p\left( \theta |\mathbf{x}\right) }{\partial \theta }$; by taking the
expectation with respect to $\mathbf{x}$ of (\ref{A - ident 2 - h+}-\ref{A -
ident 2 - h-}), one gets the following inequality:%
\begin{equation}
\lim\limits_{h\rightarrow 0}E_{\mathbf{x},\theta }\left[ \left( \frac{\psi
_{q}^{h}\left( \mathbf{x},\theta \right) }{h}\right) ^{2}\right] \leq E_{%
\mathbf{x},\theta }\left[ \left( \frac{\frac{\partial t\left( \mathbf{x}%
,\theta \right) }{\partial \theta }}{p\left( \theta |\mathbf{x}\right) }%
\right) ^{2}\right] +\min \left\{
\begin{array}{l}
E_{\mathbf{x}}\left[ \frac{5}{2}v\left( \mathbf{x},a_{\mathbf{x}}\right) -%
\frac{1}{2}v\left( \mathbf{x},b_{\mathbf{x}}\right) \right] , \\
E_{\mathbf{x}}\left[ \frac{1}{2}v\left( \mathbf{x},a_{\mathbf{x}}\right) -%
\frac{5}{2}v\left( \mathbf{x},b_{\mathbf{x}}\right) \right]
\end{array}%
\right\} ,  \label{A - GBCRB - form 1}
\end{equation}%
which, combined with (\ref{A - ident 0}) lead to (\ref{BCRB - connected
subset}). \newline
It is straightforward to extend the above rationale to the general case of
bounded intervals $\mathcal{S}_{\Theta |\mathbf{x}}=\left[ a_{\mathbf{x}},b_{%
\mathbf{x}}\right[ $, $\mathcal{S}_{\Theta |\mathbf{x}}=\left] a_{\mathbf{x}%
},b_{\mathbf{x}}\right] $, $\mathcal{S}_{\Theta |\mathbf{x}}=\left] a_{%
\mathbf{x}},b_{\mathbf{x}}\right[ $, $-\infty <a_{\mathbf{x}}<b_{\mathbf{x}%
}<+\infty $, provided that : $q\left( \mathbf{x},\theta \right) $, $\frac{%
\partial t\left( \mathbf{x},\theta \right) }{\partial \theta }$, $u\left(
\mathbf{x},{\theta }\right) $, $\frac{\partial u\left( \mathbf{x},\theta
\right) }{\partial \theta }$ and $\frac{\partial ^{2}u\left( \mathbf{x}%
,\theta \right) }{\partial ^{2}\theta }$ are bounded functions at the
vicinity of endpoints $a_{\mathbf{x}}$ and $b_{\mathbf{x}}$.\newline
Moreover, (\ref{A - GBCRB - form 1}) is uniquely integral calculus based,
thus, due to the fact that the result does not change by a finite number of
discontinuities, we conclude that the above conditions can be relaxed to : $%
g\left( \theta \right) $ and $q\left( \mathbf{x},\theta \right) p\left(
\theta |\mathbf{x}\right) $ are piecewise $\mathcal{C}^{1}$ function w.r.t. $%
\theta $ over $\mathcal{S}_{\Theta |\mathbf{x}}$.

\subsection{Case of unbounded intervals\label{A: Case of unbounded intervals}%
}

In the case of unbounded intervals, one has $a_{\mathbf{x}}=-\infty $ and/or
$b_{\mathbf{x}}=+\infty $. Then, let us define a sequence of intervals given
by
\end{subequations}
\begin{equation*}
\mathcal{S}_{\Theta |\mathbf{x}}^{l}=\left] a_{\mathbf{x}}^{l},b_{\mathbf{x}%
}^{l}\right[ ,-\infty <a_{\mathbf{x}}^{l}<b_{\mathbf{x}}^{l}<\infty \text{
subject to }\mathcal{S}_{\Theta |\mathbf{x}}^{l}\subset \mathcal{S}_{\Theta |%
\mathbf{x}}\text{ and }\lim\limits_{l\rightarrow \infty }\mathcal{S}_{\Theta
|\mathbf{x}}^{l}=\mathcal{S}_{\Theta |\mathbf{x}}
\end{equation*}%
In the same way, we define $\mathcal{S}_{\mathcal{X},\Theta }^{l}=\left\{
\left( \mathbf{x},\theta \right) ~|~\mathbf{x}\in \mathcal{S}_{\mathcal{X}}%
\text{ and }\theta \in \mathcal{S}_{\Theta |\mathbf{x}}^{l}\right\} $ and
denote:
\begin{subequations}
\begin{equation}
p^{l}\left( \mathbf{x},\theta \right) =\frac{p\left( \mathbf{x},\theta
\right) }{\diint\limits_{\mathcal{S}_{\mathcal{X},\Theta }^{l}}p\left(
\mathbf{x},\theta \right) d\mathbf{x}d\theta },\text{\quad }p^{l}\left(
\theta |\mathbf{x}\right) =\frac{p\left( \theta |\mathbf{x}\right) }{%
\dint\limits_{\mathcal{S}_{\Theta |\mathbf{x}}^{l}}p\left( \theta |\mathbf{x}%
\right) d\theta },
\end{equation}%
which defines:
\begin{equation}
E_{\mathbf{x},\theta }^{l}\left[ q\left( \mathbf{x},\theta \right) \right]
=\diint\limits_{\mathcal{S}_{\mathcal{X},\Theta }^{l}}q\left( \mathbf{x}%
,\theta \right) p^{l}\left( \mathbf{x},\theta \right) d\mathbf{x}d\theta ,%
\text{\quad }E_{\theta |\mathbf{x}}^{l}\left[ q\left( \mathbf{x},\theta
\right) \right] =\dint\limits_{\mathcal{S}_{\Theta |\mathbf{x}}^{l}}q\left(
\mathbf{x},\theta \right) p^{l}\left( \theta |\mathbf{x}\right) d\theta
\end{equation}%
Then, the analysis and results given in the previous Section \ref{A: Case of
bounded intervals} can be applied to the restricted intervals $\mathcal{S}_{%
\mathcal{X},\Theta }^{l},\mathcal{S}_{\Theta |\mathbf{x}}^{l}$ and their
associated pdfs $p^{l}\left( \mathbf{x},\theta \right) ,p^{l}\left( \theta |%
\mathbf{x}\right) $, respectively. By definition:
\end{subequations}
\begin{subequations}
\begin{equation}
\diint\limits_{\mathcal{S}_{\mathcal{X},\Theta }}\left( E_{\theta |\mathbf{x}%
}\left[ g\left( \theta \right) \right] -g\left( \theta \right) \right)
^{2}p^{l}(\mathbf{x},\theta )d\theta d\mathbf{x}\geq \diint\limits_{\mathcal{%
S}_{\mathcal{X},\Theta }^{l}}\left( E_{\theta |\mathbf{x}}\left[ g\left(
\theta \right) \right] -g\left( \theta \right) \right) ^{2}p^{l}(\mathbf{x}%
,\theta )d\theta d\mathbf{x},
\end{equation}%
that is:
\begin{equation}
\frac{E_{\mathbf{x},\theta }\left[ \left( E_{\theta |\mathbf{x}}\left[
g\left( \theta \right) \right] -g\left( \theta \right) \right) ^{2}\right] }{%
\diint\limits_{\mathcal{S}^{l}}p\left( \mathbf{x},\theta \right) d\mathbf{x}%
d\theta }\geq E_{\mathbf{x},\theta }^{l}\left[ \left( E_{\theta |\mathbf{x}}%
\left[ g\left( \theta \right) \right] -g\left( \theta \right) \right) ^{2}%
\right] .
\end{equation}%
Moreover, as:
\begin{equation}
E_{\mathbf{x},\theta }^{l}\left[ \left( E_{\theta |\mathbf{x}}\left[ g\left(
\theta \right) \right] -g\left( \theta \right) \right) ^{2}\right] \geq E_{%
\mathbf{x},\theta }^{l}\left[ \left( E_{\theta |\mathbf{x}}^{l}\left[
g\left( \theta \right) \right] -g\left( \theta \right) \right) ^{2}\right] ,
\end{equation}%
therefore:
\begin{equation}
\frac{E_{\mathbf{x},\theta }\left[ \left( E_{\theta |\mathbf{x}}\left[
g\left( \theta \right) \right] -g\left( \theta \right) \right) ^{2}\right] }{%
\diint\limits_{\mathcal{S}_{\mathcal{X},\Theta }^{l}}p\left( \mathbf{x}%
,\theta \right) d\mathbf{x}d\theta }\geq E_{\mathbf{x},\theta }^{l}\left[
\left( E_{\theta |\mathbf{x}}^{l}\left[ g\left( \theta \right) \right]
-g\left( \theta \right) \right) ^{2}\right] .
\end{equation}%
Finally, since $\lim\limits_{l\rightarrow \infty }\mathcal{S}_{\Theta |%
\mathbf{x}}^{l}=\mathcal{S}_{\Theta |\mathbf{x}}$, one has:
\begin{equation}
E_{\mathbf{x},\theta }\left[ \left( E_{\theta |\mathbf{x}}\left[ g\left(
\theta \right) \right] -g\left( \theta \right) \right) ^{2}\right] \geq
\lim\limits_{l\rightarrow \infty }E_{\mathbf{x},\theta }^{l}\left[ \left(
E_{\theta |\mathbf{x}}^{l}\left[ g\left( \theta \right) \right] -g\left(
\theta \right) \right) ^{2}\right] ,
\end{equation}%
which allows to state Proposition 1 as the limiting form of the bounded
intervals case.

\subsection{Case of a countable union of disjoint intervals of $%
\mathbb{R}
$\label{A: Case of a countable union of disjoint intervals}}

First we consider the case where $\mathcal{S}_{\Theta |\mathbf{x}}$ results
from a finite union of disjoint bounded intervals $\mathcal{I}_{\Theta |%
\mathbf{x}}^{k}\varsubsetneq
\mathbb{R}
$:
\end{subequations}
\begin{equation*}
\forall \mathbf{x}\in \mathcal{S}_{\mathcal{X}},\quad \mathcal{S}_{\Theta |%
\mathbf{x}}=\tbigcup\limits_{1\leq k\leq K_{\mathbf{x}}}\mathcal{I}_{\Theta |%
\mathbf{x}}^{k},\text{ such that\ }K_{\mathbf{x}}\in \mathbb{N}\text{ and }%
\mathcal{I}_{\Theta |\mathbf{x}}^{k}\cap \mathcal{I}_{\Theta |\mathbf{x}%
}^{l}=\varnothing \text{ if }k\neq l.
\end{equation*}%
Let us denote the endpoints of $\mathcal{I}_{\Theta |\mathbf{x}}^{k}$ by $a_{%
\mathbf{x}}^{k}$ and $b_{\mathbf{x}}^{k}$, $a_{\mathbf{x}}^{k}<a_{\mathbf{x}%
}^{k}$. Then%
\begin{equation}
\left\{
\begin{array}{l}
E_{\theta |\mathbf{x}}\left[ g\left( \theta \right) \frac{\psi
_{q}^{h}\left( \mathbf{x},\theta \right) }{h}\right] =-\tsum%
\limits_{k=1}^{K_{\mathbf{x}}}\dint\limits_{a_{\mathbf{x}}^{k}+h}^{b_{%
\mathbf{x}}^{k}}\left( \frac{g\left( \theta -h\right) -g\left( \theta
\right) }{-h}\right) q\left( \mathbf{x},\theta \right) p\left( \theta |%
\mathbf{x}\right) d\theta , \\
E_{\theta |\mathbf{x}}\left[ g\left( \theta \right) \frac{\psi
_{q}^{h}\left( \mathbf{x},\theta \right) }{h}\right] =-\tsum%
\limits_{k=1}^{K_{\mathbf{x}}}\dint\limits_{a_{\mathbf{x}}^{k}}^{b_{\mathbf{x%
}}^{k}-h}\left( \frac{g\left( \theta +h\right) -g\left( \theta \right) }{h}%
\right) q\left( \mathbf{x},\theta \right) p\left( \theta |\mathbf{x}\right)
d\theta , \\
E_{\theta |\mathbf{x}}\left[ \frac{\psi _{q}^{h}\left( \mathbf{x},\theta
\right) ^{2}}{h^{2}}\right] =\frac{1}{h^{2}}\tsum\limits_{k=1}^{K_{\mathbf{x}%
}}\dint\limits_{a_{\mathbf{x}}^{k}+h}^{b_{\mathbf{x}}^{k}}\left(
\begin{array}{l}
q\left( \mathbf{x},\theta +h\right) p\left( \theta +h|\mathbf{x}\right) 1_{%
\mathcal{S}_{\Theta |\mathbf{x}}}\left( \theta +h\right) \\
-q\left( \mathbf{x},\theta \right) p\left( \theta |\mathbf{x}\right) 1_{%
\mathcal{S}_{\Theta |\mathbf{x}}}\left( \theta -h\right)%
\end{array}%
\right) ^{2}\frac{1}{p\left( \theta |\mathbf{x}\right) }d\theta , \\
E_{\theta |\mathbf{x}}\left[ \frac{\psi _{q}^{-h}\left( \mathbf{x},\theta
\right) ^{2}}{\left( -h\right) ^{2}}\right] =\frac{1}{h^{2}}%
\tsum\limits_{k=1}^{K_{\mathbf{x}}}\dint\limits_{a_{\mathbf{x}}^{k}}^{b_{%
\mathbf{x}}^{k}-h}\left(
\begin{array}{l}
q\left( \mathbf{x},\theta -h\right) p\left( \theta -h|\mathbf{x}\right) 1_{%
\mathcal{S}_{\Theta |\mathbf{x}}}\left( \theta -h\right) \\
-q\left( \mathbf{x},\theta \right) p\left( \theta |\mathbf{x}\right) 1_{%
\mathcal{S}_{\Theta |\mathbf{x}}}\left( \theta +h\right)%
\end{array}%
\right) ^{2}\frac{1}{p\left( \theta |\mathbf{x}\right) }d\theta ,%
\end{array}%
\right.  \label{A - eq 10}
\end{equation}%
which means that all the rationale introduced in Appendix \ref{A: Case of
bounded intervals} can be applied to each $\mathcal{I}_{\Theta |\mathbf{x}%
}^{k}$ individually. Therefore if, $1\leq k\leq K_{\mathbf{x}}$:\newline
$\bullet ~q\left( \mathbf{x},\theta \right) $ admits a finite limit at
endpoints of $\mathcal{I}_{\Theta |\mathbf{x}}^{k}$,\newline
$\bullet ~g\left( \theta \right) $ is piecewise $\mathcal{C}^{1}$ w.r.t. $%
\theta $ over $\mathcal{I}_{\Theta |\mathbf{x}}^{k}$,\newline
$\bullet ~t\left( \mathbf{x},\theta \right) \triangleq q\left( \mathbf{x}%
,\theta \right) p\left( \theta |\mathbf{x}\right) $ is piecewise $\mathcal{C}%
^{1}$ w.r.t. $\theta $ over $\mathcal{I}_{\Theta |\mathbf{x}}^{k}$ and such
as $\frac{\partial t\left( \mathbf{x},\theta \right) }{\partial \theta }$
admits a finite limit at endpoints of $\mathcal{I}_{\Theta |\mathbf{x}}^{k}$,%
\newline
${\small \bullet }~u\left( \mathbf{x},\theta \right) \triangleq q\left(
\mathbf{x},\theta \right) ^{2}p\left( \theta |\mathbf{x}\right) $ is $%
\mathcal{C}^{2}$ w.r.t. $\theta $ at the vicinity of endpoints of $\mathcal{I%
}_{\Theta |\mathbf{x}}^{k}$ and such as $u\left( \mathbf{x},\theta \right) ,%
\frac{\partial u\left( \mathbf{x},\theta \right) }{\partial \theta }$ and $%
\frac{\partial ^{2}u\left( \mathbf{x},\theta \right) }{\partial ^{2}\theta }$
admit a finite limit at endpoints of $\mathcal{I}_{\Theta |\mathbf{x}}^{k}$,%
\newline
then a necessary and sufficient condition in order to obtain a non trivial $%
\mathrm{BCRB}_{q}$ is:
\begin{subequations}
\begin{equation}
q\left( \mathbf{x},a_{\mathbf{x}}^{k}\right) p\left( a_{\mathbf{x}}^{k}|%
\mathbf{x}\right) =q\left( \mathbf{x},b_{\mathbf{x}}^{k}\right) p\left( b_{%
\mathbf{x}}^{k}|\mathbf{x}\right) =0,\quad 1\leq k\leq K_{\mathbf{x}},
\end{equation}%
leading to:%
\begin{equation}
\mathrm{BCRB}_{q}\left( g\left( \theta \right) \right) =\frac{E_{\mathbf{x}%
,\theta }\left[ \frac{dg\left( \theta \right) }{d\theta }q\left( \mathbf{x}%
,\theta \right) \right] ^{2}}{E_{\mathbf{x},\theta }\left[ \left( \frac{%
\frac{\partial t\left( \mathbf{x},\theta \right) }{\partial \theta }}{%
p\left( \theta |\mathbf{x}\right) }\right) ^{2}\right] +\min \left\{
\begin{array}{l}
\tsum\limits_{k=1}^{K_{\mathbf{x}}}E_{\mathbf{x}}\left[ \frac{5}{2}v\left(
\mathbf{x},a_{\mathbf{x}}^{k}\right) -\frac{1}{2}v\left( \mathbf{x},b_{%
\mathbf{x}}^{k}\right) \right] , \\
\tsum\limits_{k=1}^{K_{\mathbf{x}}}E_{\mathbf{x}}\left[ \frac{1}{2}v\left(
\mathbf{x},a_{\mathbf{x}}^{k}\right) -\frac{5}{2}v\left( \mathbf{x},b_{%
\mathbf{x}}^{k}\right) \right]%
\end{array}%
\right\} }
\end{equation}%
If $\mathcal{I}_{\Theta |\mathbf{x}}^{1}$ is a left-unbounded interval
and/or $\mathcal{I}_{\Theta |\mathbf{x}}^{K_{\mathbf{x}}}$ is a
right-unbounded interval, then the above results still hold provided that
(see Appendix \ref{A: Case of unbounded intervals}) for $k=1$ and/or $k=K_{%
\mathbf{x}}:$%
\end{subequations}
\begin{subequations}
\begin{eqnarray}
q\left( \mathbf{x},a_{\mathbf{x}}^{k}\right) p\left( a_{\mathbf{x}}^{k}|%
\mathbf{x}\right) &\triangleq &\lim\limits_{\theta \rightarrow a_{\mathbf{x}%
}^{k}}q\left( \mathbf{x},\theta \right) p\left( \theta \right) ,\quad \quad
\quad \quad ~q\left( \mathbf{x},b_{\mathbf{x}}^{k}\right) p\left( b_{\mathbf{%
x}}^{k}|\mathbf{x}\right) \triangleq \lim\limits_{\theta \rightarrow b_{%
\mathbf{x}}^{k}}q\left( \mathbf{x},\theta \right) p\left( \theta \right) , \\
q\left( \mathbf{x},a_{\mathbf{x}}^{k}\right) ^{2}\frac{\partial p\left( a_{%
\mathbf{x}}^{k}|\mathbf{x}\right) }{\partial \theta } &\triangleq
&\lim\limits_{\theta \rightarrow a_{\mathbf{x}}^{k}}q\left( \mathbf{x}%
,\theta \right) ^{2}\frac{\partial p\left( \theta |\mathbf{x}\right) }{%
\partial \theta },~q\left( \mathbf{x},b_{\mathbf{x}}^{k}\right) ^{2}\frac{%
\partial p\left( b_{\mathbf{x}}^{k}|\mathbf{x}\right) }{\partial \theta }%
\triangleq \lim\limits_{\theta \rightarrow b_{\mathbf{x}}^{k}}q\left(
\mathbf{x},\theta \right) ^{2}\frac{\partial p\left( \theta |\mathbf{x}%
\right) }{\partial \theta }.\quad \quad \quad
\end{eqnarray}%
Last, (\ref{A - eq 10}) holds as well for a a countable union of disjoint
intervals of $%
\mathbb{R}
$, QED.

\subsection{Proof of (\protect\ref{eq_annexe})\label{A: limite 1}}

Let us consider a function $u:\mathcal{X}\times \mathcal{[}a,b\mathcal{]}%
\rightarrow \mathbb{R}$, $a<b,$ in which $u$ is of $\mathcal{C}^{2}$ w.r.t. $%
\theta $. Then, $\forall \mathbf{x}\in \mathcal{S}_{\mathcal{X}}$:
\end{subequations}
\begin{subequations}
\begin{eqnarray}
\dint\limits_{a}^{b}u\left( \mathbf{x},\theta \right) d\theta &=&\left(
b-a\right) u\left( \mathbf{x},a\right) +\frac{1}{2}\left( b-a\right) ^{2}%
\frac{\partial u\left( \mathbf{x},a\right) }{\partial \theta }+\frac{1}{2}%
\dint\limits_{a}^{b}\left( b-\theta \right) ^{2}\frac{\partial ^{2}u\left(
\mathbf{x},\theta \right) }{\partial ^{2}\theta }d\theta , \\
\dint\limits_{a}^{b}u\left( \mathbf{x},\theta \right) d\theta &=&\left(
b-a\right) u\left( \mathbf{x},b\right) -\frac{1}{2}\left( b-a\right) ^{2}%
\frac{\partial u\left( \mathbf{x},b\right) }{\partial \theta }+\frac{1}{2}%
\dint\limits_{a}^{b}\left( \theta -a\right) ^{2}\frac{\partial ^{2}u\left(
\mathbf{x},\theta \right) }{\partial ^{2}\theta }d\theta .\quad
\end{eqnarray}%
Consequently, if $u$ is of $\mathcal{C}^{2}$ w.r.t. $\theta $ over $\mathcal{%
X}\times \mathcal{[}a_{\mathbf{x}},a_{\mathbf{x}}+h\mathcal{]}$ and $%
\mathcal{X}\times \mathcal{[}b_{\mathbf{x}}-h,b_{\mathbf{x}}\mathcal{]}$, $%
0<h$, $a_{\mathbf{x}}<b_{\mathbf{x}}$, one obtains:
\end{subequations}
\begin{subequations}
\begin{eqnarray}
\dint\limits_{a_{\mathbf{x}}}^{a_{\mathbf{x}}+h}u\left( \mathbf{x},\theta
\right) d\theta &=&hu\left( \mathbf{x},a_{\mathbf{x}}\right) +\frac{h^{2}}{2}%
\left( \frac{\partial u\left( \mathbf{x},a_{\mathbf{x}}\right) }{\partial
\theta }+\dint\limits_{a_{\mathbf{x}}}^{a_{\mathbf{x}}+h}\left( \frac{\theta
-a_{\mathbf{x}}-h}{h}\right) ^{2}\frac{\partial ^{2}u\left( \mathbf{x}%
,\theta \right) }{\partial \theta ^{2}}d\theta \right) , \\
\dint\limits_{b_{\mathbf{x}}-h}^{b_{\mathbf{x}}}u\left( \mathbf{x},\theta
\right) d\theta &=&hu\left( \mathbf{x},b_{\mathbf{x}}\right) -\frac{h^{2}}{2}%
\left( \frac{\partial u\left( \mathbf{x},b_{\mathbf{x}}\right) }{\partial
\theta }+\dint\limits_{b_{\mathbf{x}}-h}^{b_{\mathbf{x}}}\left( \frac{b_{%
\mathbf{x}}-\theta -h}{h}\right) ^{2}\frac{\partial ^{2}u\left( \mathbf{x}%
,\theta \right) }{\partial \theta ^{2}}d\theta \right) .\qquad
\end{eqnarray}%
Additionally, as:
\end{subequations}
\begin{subequations}
\begin{equation}
\left\{
\begin{array}{r}
a_{\mathbf{x}}\leq \theta \leq a_{\mathbf{x}}+h\Rightarrow \left( \frac{%
\theta -a_{\mathbf{x}}-h}{h}\right) ^{2}\leq 1 \\
b_{\mathbf{x}}-h\leq \theta \leq b_{\mathbf{x}}\quad \quad \Rightarrow
\left( \frac{b_{\mathbf{x}}-\theta -h}{h}\right) ^{2}\leq 1%
\end{array}%
\right.
\end{equation}%
therefore:
\begin{eqnarray}
\left\vert \dint\limits_{a_{\mathbf{x}}}^{a_{\mathbf{x}}+h}\left( \frac{%
\theta -a_{\mathbf{x}}-h}{h}\right) ^{2}\frac{\partial ^{2}u\left( \mathbf{x}%
,\theta \right) }{\partial \theta ^{2}}d\theta \right\vert &\leq
&\dint\limits_{a_{\mathbf{x}}}^{a_{\mathbf{x}}+h}\left( \frac{\theta -a_{%
\mathbf{x}}-h}{h}\right) ^{2}\left\vert \frac{\partial ^{2}u\left( \mathbf{x}%
,\theta \right) }{\partial \theta ^{2}}\right\vert d\theta \leq
\dint\limits_{a_{\mathbf{x}}}^{a_{\mathbf{x}}+h}\left\vert \frac{\partial
^{2}u\left( \mathbf{x},\theta \right) }{\partial \theta ^{2}}\right\vert
d\theta , \\
\left\vert \dint\limits_{b_{\mathbf{x}}-h}^{b_{\mathbf{x}}}\left( \frac{b_{%
\mathbf{x}}-\theta -h}{h}\right) ^{2}\frac{\partial ^{2}u\left( \mathbf{x}%
,\theta \right) }{\partial \theta ^{2}}d\theta \right\vert &\leq
&\dint\limits_{b_{\mathbf{x}}-h}^{b_{\mathbf{x}}}\left( \frac{b_{\mathbf{x}%
}-\theta -h}{h}\right) ^{2}\left\vert \frac{\partial ^{2}u\left( \mathbf{x}%
,\theta \right) }{\partial \theta ^{2}}\right\vert d\theta \leq
\dint\limits_{b_{\mathbf{x}}-h}^{b_{\mathbf{x}}}\left\vert \frac{\partial
^{2}u\left( \mathbf{x},\theta \right) }{\partial \theta ^{2}}\right\vert
d\theta .\qquad \quad
\end{eqnarray}%
Finally, since $u$ is of $\mathcal{C}^{2}$ w.r.t. $\theta $ over $\mathcal{X}%
\times \mathcal{[}a_{\mathbf{x}},a_{\mathbf{x}}+h\mathcal{]}$ and $\mathcal{X%
}\times \mathcal{[}b_{\mathbf{x}}-h,b_{\mathbf{x}}\mathcal{]}$, then $%
\left\vert \frac{\partial ^{2}u\left( \mathbf{x},a_{\mathbf{x}}\right) }{%
\partial \theta ^{2}}\right\vert $ and $\left\vert \frac{\partial
^{2}u\left( \mathbf{x},b_{\mathbf{x}}\right) }{\partial \theta ^{2}}%
\right\vert $ are finite values and:
\end{subequations}
\begin{multline}
\lim_{h\rightarrow 0^{+}}\frac{1}{h^{2}}\left( \dint\limits_{a_{\mathbf{x}%
}}^{a_{\mathbf{x}}+h}u\left( \mathbf{x},\theta \right) d\theta
+\dint\limits_{b_{\mathbf{x}}-h}^{b_{\mathbf{x}}}u\left( \mathbf{x},\theta
\right) d\theta \right) =\frac{u\left( \mathbf{x},b_{\mathbf{x}}\right)
+u\left( \mathbf{x},a_{\mathbf{x}}\right) }{h}  \label{limite 1} \\
+\frac{1}{2}\left( \frac{\partial u\left( \mathbf{x},a_{\mathbf{x}}\right) }{%
\partial \theta }-\frac{\partial u\left( \mathbf{x},b_{\mathbf{x}}\right) }{%
\partial \theta }\right)
\end{multline}

\end{document}